\documentclass[preprint,12pt]{elsarticle}

\usepackage{lmodern}

\usepackage[version=4]{mhchem}

\usepackage[english]{babel}
\usepackage[utf8]{inputenc}
\usepackage[T1]{fontenc}

\usepackage{sidecap}
\usepackage{wasysym}
\usepackage[makeroom]{cancel}
\usepackage[numbers]{natbib}
\biboptions{sort&compress}
\usepackage{graphicx}
\usepackage{epstopdf}
\usepackage{rotating}
\usepackage{wrapfig}
\usepackage{float}
\usepackage{subfigure}
\usepackage{color}
\usepackage{hyperref}
\usepackage{datetime}
\usepackage{amsmath,amssymb,amsthm}
\usepackage{amsfonts}
\usepackage{mathtools}
\usepackage{cases}
\usepackage{array}
\usepackage[binary-units=true]{siunitx}
\usepackage{upgreek}

\usepackage[title]{appendix}

\usepackage[stable]{footmisc}

\usepackage{booktabs}
\usepackage{multirow}
\usepackage{multicol}
\usepackage{caption}

\usepackage{bm}

\usepackage{outlines}
\usepackage{dirtree}
\usepackage[dvipsnames]{xcolor}
\usepackage{framed}
\usepackage{color}
\usepackage{emptypage}
\usepackage{verbatim} 
\usepackage{textcomp}
\usepackage[section]{placeins}
\usepackage{pdfpages}
\usepackage{enumitem}
\setlist[enumerate]{itemsep=0mm}
\usepackage[flushleft]{threeparttable}
\usepackage{array,booktabs,makecell}
\usepackage{tcolorbox}
\usepackage{xargs}
\usepackage{tikz}
\usetikzlibrary{shapes}
\usepackage[textsize = tiny, color = blue!30]{todonotes}
\usepackage{marginnote}
\usepackage{lineno}
\usepackage[top=3.5cm, bottom=3.5cm, outer=3.5cm, inner=3.5cm, heightrounded, marginparwidth=3.cm, marginparsep=0.5cm]{geometry}

\journal{Journal}

\newcommand{\subrm}[1]{{_{\mathrm{#1} } }} 
\newcommand{\subsuprm}[2]{{ _{\mathrm{#1}}^{\mathrm{#2} } }} 
\newcommand{\cm}[0]{{~\si{\centi\metre}}} 
\newcommand{\mm}[0]{{~\si{\milli\metre}}} 
\newcommand{\um}[0]{{~\si{\micro\metre}}}  
\newcommand{\nm}[0]{{~\si{\nano\metre}}} 
\newcommand{\IRMband}[2]{#1\textrm{-}\SI{#2}{\um}}
\newcommand{\taustarext}[1][]{ \hat{\tau}^{\textrm{ext}}_{\Delta\lambda_{#1}} }
\newcommand{\epsstar}[1][]{\hat{\varepsilon}_{\Delta\lambda_{#1}}}
\newcommand{\Dt}{\Delta t}
\newcommand{\Ts}{T\subrm{s}}
\newcommand{\us}{u\subrm{s}}
\newcommand{\hs}{h\subrm{s}}
\newcommand{\deltastar}{\delta^{*}}
\newcommand{\Tstag}{T_{\mathrm{stag}}} 		
\newcommand{\Ttc}{T_{\mathrm{TC}}} 		
\newcommand{\pc}{p_{\mathrm{c}}} 			
\newcommand{\pdyn}{p_{\mathrm{dyn}}} 		
\newcommand{\qcw}{\dot{q}_{\mathrm{cw}}} 	
\newcommand{\mdot}{\dot{m}_{\mathrm{gas}}} 	
\newcommand{\Pel}{P_{\mathrm{el}}}	 		

\begin{document}

\sloppy

\begin{frontmatter}

\title{Emissivity of oxidizing titanium in simulated atmospheric entry flows}

\author[inst1,inst2, cor1]{Andrea Fagnani}
\ead{a.fagnani11@gmail.com}
\cortext[cor1]{Corresponding author}
\author[inst1]{Bernd Helber}
\ead{bernd.helber@vki.ac.be}
\author[inst2]{Annick Hubin}
\ead{annick.hubin@vub.be}
\author[inst1]{Olivier Chazot}
\ead{olivier.chazot@vki.ac.be}

\affiliation[inst1]{organization={Aeronautics and Aerospace Department, von Karman Institute for Fluid Dynamics},
            addressline={Chaussée de Waterloo 72}, 
            city={Rhode-st-Genèse},
            postcode={1640}, 
            country={Belgium}}

\affiliation[inst2]{organization={Materials and Chemistry Department, Vrije Universiteit Brussel},
	addressline={Plainlaan~2}, 
	city={Brussel},
	postcode={1150}, 
	country={Belgium}}

\begin{abstract}
    The aerothermal demise of titanium components plays a critical role in the uncontrolled re-entry of space debris from low-Earth orbit. Exposure to high temperatures and dissociated oxygen environments promotes rapid oxidation, significantly influencing the material degradation and surface thermal balance. This study presents time-resolved infrared emissivity measurements of both Grade 2 and Grade 5 titanium samples across five wavelength bands during exposure to entry-relevant conditions simulated in the Plasmatron facility at the von Karman Institute for Fluid Dynamics. The results reveal a dynamic evolution of emissivity throughout the test, including a pronounced drop associated with a characteristic surface temperature jump that is not captured in existing literature data. Post-test electron microscopy highlights a diverse oxide layer morphology at the microscale. Although plasma wind tunnel experiments reproduce only a subset of flight-relevant phenomena to space-debris entry, these findings demonstrate that the complex coupling between oxidation and surface radiative behavior is not adequately captured by conventional pre- and post-test analyses, highlighting their limitations in resolving in-situ emissivity evolution.
\end{abstract}


\end{frontmatter}


%
\section{Introduction}
\label{sec:OES:intro}
\graphicspath{{figures/}}

The increasing population of space debris poses a significant threat to the sustainability of future space activities \cite{klinkrad2006, esaspacedebrisoffice2025}. Current mitigation guidelines therefore mandate the safe post-mission disposal of spacecraft operating within protected orbital regions \cite{europeanspaceagency2023}. Among the available disposal strategies, the uncontrolled atmospheric re-entry remains a cost-effective option for Low Earth Orbit (LEO) missions \cite{Waswa2013}. Within this framework, the Design for Demise (D4D) philosophy seeks to ensure the complete disintegration of spacecraft components during re-entry, thereby minimizing the likelihood of surviving debris reaching the ground and reducing the associated risks to people and property \cite{Lemmens2015}.

Propellant tanks, typically made of high strength metallic liners
and possibly reinforced with carbon fiber overwrap, represent the majority of debris recovered on ground so far \cite{ESA_space_debris_recoveries}.
Grade 2 (commercially pure) and Grade 5 (Ti-6Al-4V) titanium alloys are widely employed materials for such components, for which the high melting temperature, relatively high density and heat capacity, contribute to their survivability under LEO uncontrolled entries  \cite{Lips2017}. 
In addition, titanium and its alloy elements are highly reactive to oxygen. A various and complex behavior of titanium oxidation is found in the literature, depending on the specific thermodynamic and atmospheric conditions.  Oxidation significantly affects the material surface properties, in particular its emissivity, thus impacting the surface thermal balance and resulting aerothermal degradation  \cite{Prevereaud2016}.

A thorough understanding of the oxidation impact on the emissivity under atmospheric entry conditions is therefore critical for accurately assessing debris survivability during re-entry.
Previous works have characterized the change in emissivity of oxidized samples during or after exposure to dissociated oxygen atmosphere \cite{Prevereaud2016, barka2018, Balat-Pichelin2020, Balat-Pichelin2020a}, or after exposure to plasma wind tunnel flows \cite{Bonvoisin2023, pagan2016}. 
This paper advances on existing literature by reporting time-resolved in situ emissivity measurements during oxidation of titanium samples exposed to simulated re-entry conditions in the VKI Plasmatron facility. The test conditions provide a closer reproduction of the hypersonic boundary layer with respect to previously reported emissivity studies.
An improved free-stream flow characterization methodology, recently presented in Ref.~\cite{fagnani2026}, was used to select the experimental conditions capable of reproducing enthalpy levels and flight radii close to those of an actual trajectory point during flight. 
The technique presented in Ref.~\cite{fagnani2024} was used for in situ directional emissivity measurements, highlighting the dynamic evolution as oxidation occurs.
The spatial calibration of camera frames, initially developed for IR thermography in Ref.~\cite{fagnani2023a}, was adapted to detail the spatial evolution of the oxide crust during the experiment from visual range camera frames.
The in situ analysis is complemented with post-test Scanning Electron Microscopy (SEM) and Energy Dispersive x-ray Spectroscopy (EDS), providing evidence of the diverse morphology and microstructure of the oxide scales developed in the oxygen rich environment.

Results show a dynamic evolution of emissivity across several infrared bands, with a steady increase during oxidation followed by a sharp drop coinciding with a temperature jump, occurring between 1500 K and 1670 K. Comparison with existing emissivity data highlights relevant deviations, indicating that available correlations may fail to reproduce the observed behavior.
These findings reveal the complex coupling between oxidation, melting, and radiative properties of titanium under high-enthalpy conditions, underscoring the need for refined models in spacecraft demise and atmospheric entry simulations.

%
\section{Oxidation and emissivity behavior of titanium}

Titanium has a high affinity for oxygen and nitrogen. Under atmospheric conditions, it readily forms \ce{TiO2}, predominantly in its stable rutile phase \cite{Gyorgy2007}. 
Alloying elements influence its oxidation behavior, with aluminum typically forming a stable and protective $\alpha$-\ce{Al2O3} layer, and vanadium degrading its high-temperature resistance \cite{Shida1996}.
Exposure to oxygen-rich atmospheres produces an oxide layer and an underlying oxygen diffusion zone \cite{Guleryuz2009}. The oxide grows slowly at room temperature but thickens above $\SI{200}{\degreeCelsius}$, with diffusion becoming significant beyond $\SI{400}{\degreeCelsius}$. Oxidation accelerates markedly above $\SI{600}{\degreeCelsius}$, leading to non-protective scales that spall and allow deep oxygen penetration beyond $\SI{800}{\degreeCelsius}$.

\citet{Garbacz2003} reported a multi-layer oxide where $\ce{Al2O3}$ dominated the outer region, and $\ce{V2O5}$ concentration increased towards the surface. Cracking and delamination were linked to the $\ce{TiO}\rightarrow \ce{TiO2}$ transformation, generating stresses that degrade adhesion and modify oxidation kinetics from parabolic to linear above $\SI{700}{\degreeCelsius}$. Similar high-temperature behavior was observed by \citet{rajabi2020}, who found the oxide evolving from compact to porous and spalling above $\SI{1000}{\degreeCelsius}$. A sudden oxidation surge occurred near a radiance temperature of $\SI{1520}{\degreeCelsius}$, and was attributed to exothermic reactions and potential melting. The onset of oxidation was also found to correlate positively with the heating rate.

Although titanium oxidation in atmospheric air is well characterized, such results poorly represent atmospheric entry conditions. The former typically involve relatively low temperatures and long oxidation times, while the latter are characterized by a short-duration, high-temperature exposure with abundance of atomic oxygen due to dissociation in the shock layer.
Studies motivated by the Columbia accident \cite{buckner2016, ochoa2021} showed the rapid, exothermic reactions of TA6V at high temperature, preventing the formation of protective layers. 
More recent experiments in the MESOX facility \cite{Prevereaud2016, barka2018} simulated dissociated air with up to 80\% in atomic oxygen, revealing enhanced delamination compared to standard air. Short exposures ($<\SI{100}{\second}$) produced adherent oxides with parabolic kinetics, while longer ones led to spallation and linear kinetics. The oxidation rate increased with temperature but depended weakly on pressure (300–2000 Pa) \cite{Balat-Pichelin2020}. Microscopy revealed brittle, multilayer rutile scales with residual $\ce{TiO}$ and $\ce{Ti2O3}$ near the interface.
Recent plasma wind tunnel tests at CORIA’s SOUPLIN facility \cite{brault2025} confirmed non-protective oxidation under entry-representative enthalpies (4.6–7.3 MJ/kg) and pressures (1400–4500 Pa). Rutile dominated the oxide, with $\ce{Al2O3}$ and $\ce{Al2TiO5}$ forming above $\SI{1150}{\degreeCelsius}$. Extensive spallation, compressive stresses, and diffusion-limited kinetics were observed, consistent with earlier MESOX data.
Further plasma wind tunnel experiments and microscale analyses by \citet{brault2026} showed predominant formation of a rutile $\ce{TiO2}$ layer with a thickness on the order of $5-\SI{6}{\micro\meter}$ and revealing a sharp transition in the O/Ti ratio from near-stoichiometric $\ce{TiO2}$ at the surface to negligible oxygen content within the underlying substrate.

Accompanying the complex morphology and mechanical behavior of Ti-6Al-4V oxides at high temperatures is a significant change in emissivity.
\citet{neuer1988} showed that the emissivity evolution is governed by thin-film optical interference arising from the growing $\ce{TiO2}$ layer, and it is a function of oxide thickness rather than temperature alone.
Spectral and integrated directional emissivities of oxidizing Ti-6Al-4V were measured by \citet{elbakali2021}, showing values rising from 0.3 at $\SI{600}{\degreeCelsius}$ to 0.55 at $\SI{1000}{\degreeCelsius}$ over 1.3–$\SI{13}{\micro\meter}$, and a dependence on scale composition rather than temperature alone.
\citet{li2016} measured a gradual increase in emissivity with oxidation time below 873~K, while strong oscillations appeared above 923~K, which were attributed to optical interference effects within the growing \ce{TiO2} film. \citet{zhang2016} similarly observed an increase in emissivity due to change in composition and roughness of the oxidized titanium surface, where the oxide thickness was identified as responsible for optical interference effects.
\citet{cezairliyan1977a} measured the normal spectral emissivity up to the melting point, yielding a value of 0.395 at 653~nm. 

In the context of plasma wind tunnel experiments, \citet{pagan2016} measured the total and spectral emissivity of Ti--6Al--4V before and after re-entry representative experiments, revealing a pronounced increase up to 0.9 as rutile $\ce{TiO2}$ scales were formed.
A later study by \citet{Balat-Pichelin2020a} measured the emissivity of the Ti–6Al–4V alloy between 1000 and 1800 K, showing a strong dependence on surface state and oxidation history. As-received samples exhibited low hemispherical emissivity (0.2–0.25), while oxidized specimens reached values up to 0.8–0.9 due to the formation of $\ce{TiO2}$ scales. Using these temperature-dependent emissivity laws into the CNES DEBRISK code \cite{omaly2013} revealed a considerable impact on the predicted wall temperatures of re-entering titanium components.

Compared to previous works that have primarily characterized the emissivity of pre- and post-tested samples, or during oxidation in controlled environments, the present work aims at revealing the dynamic change in emissivity and its relation to the oxidation status during exposure to simulated atmospheric entry flows.

%
\section{Experimental methods}
\graphicspath{figures/}
\label{sec:OES:setup}

\subsection{The VKI Plasmatron facility}
\begin{figure}[h]
	\centering
	\subfigure[]
	{\includegraphics[trim={0cm, 0cm, 15cm, 0cm}, clip, width=.65\textwidth]{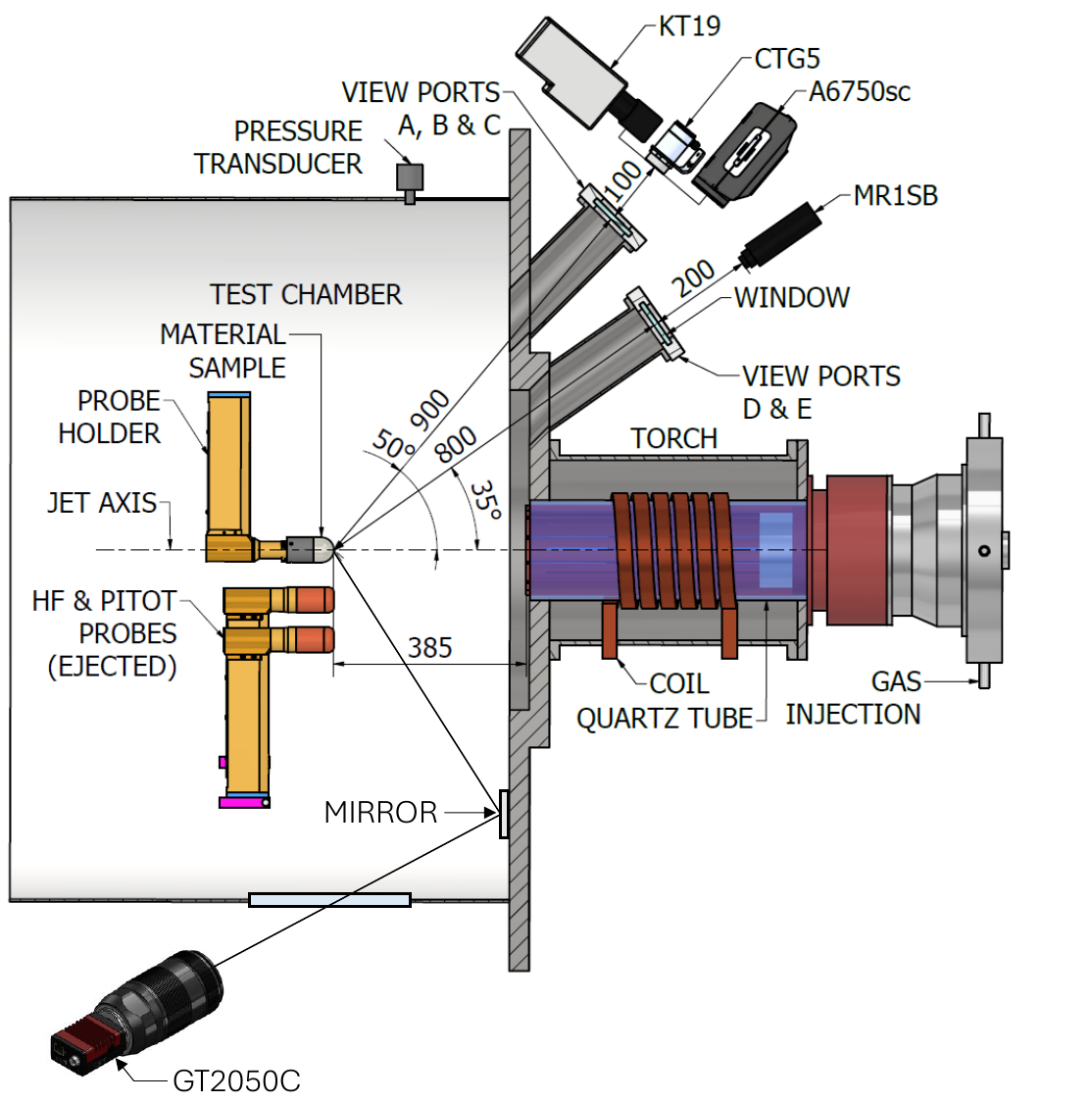}		\label{fig:figure6_3b}}\\
	\subfigure[]
	{\includegraphics[trim={0cm, 21cm, 10cm, 0cm}, clip, width=.70\textwidth]{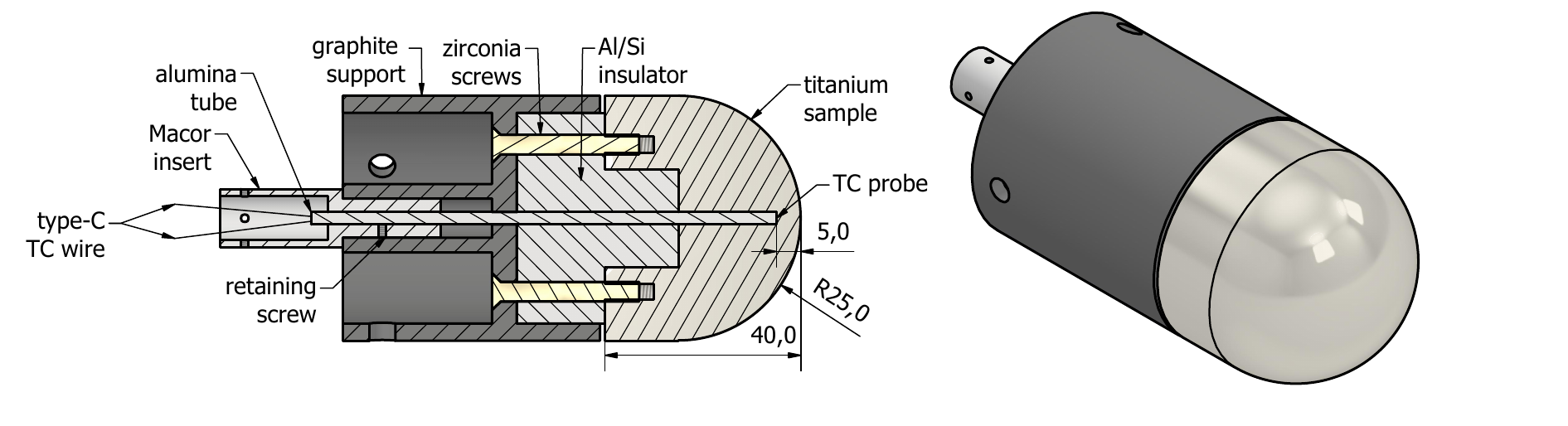}	\label{fig:figure1b}	}
	\caption[]{(a) Section view of the VKI Plasmatron facility, showing the ICP torch and the instrument location. (b) Sketch of the probe geometry employed for the present experiments. The material sample is insulated from the cooled probe holder through an alumina-silica disc and secured by means of 4 zirconia screws. A type-C thermocouple probe extends until 5~mm below the surface at stagnation point.}
	\label{fig:a}
\end{figure}

The VKI Plasmatron facility features a $ \SI{160}{\milli\meter} $ diameter ICP torch, powered by a $ \SI{400}{\kilo \hertz} $, $ \SI{1.2}{\mega \watt} $, $ \SI{2}{\kilo \volt} $ electric generator, and connected to a $ \SI{1.4}{\meter} $ diameter, $ \SI{2.4}{\meter} $ long test chamber.  An extensive description of the facility and its performance was given by Bottin~et~al.~\cite{Bottin2000}.
Fig.~\ref{fig:figure6_3b} shows a schematic section of the ICP torch, test chamber, and instrumentation set-up described in the following sections.
The torch is made up of a quartz tube, surrounded by a six-turn flat coil inductor, and supplied by a gas injection system. The electric power to the coil, $ \Pel $, is monitored by a voltage-current probe, while a calibrated flow meter (F-203AV, Bronkhorst High-Tech B.V, NL) controls the mass flow rate, $ \mdot $, of the test gas supplied to the torch with a $ \pm0.5\% $ accuracy. A synthetic air mixture, containing $ 21\% \pm 1\%$ $\ce{O2}$, less than 500~ppm of $ \ce{CO2} $ and $\ce{N2}$ by balance (OXAZ20, Btg Solgroup, BE) was used to minimize the level of contaminants in the test gas. The gas is heated by electromagnetic induction to provide a chemically pure plasma flow. Pressure in the test chamber, $ \pc $, is measured by an absolute pressure transducer (Membranovac DM 12, Leybold GmbH, DE) to an accuracy of 1\%.
Three movable holding arms can be swung into the plasma flow interchangeably by a pneumatic mechanism at a distance $z=\SI{385}{\milli\meter}$ from the torch exit. Two of these hold copper-cooled probes with 50~mm diameter hemispherical heads, used for reference cold-wall heat flux, $\qcw$, and dynamic pressure, $\pdyn$, measurements, while the third one holds the material sample under study.

\subsection{Material samples and post-test analysis}

We present experimental results obtained from three material samples, namely, two titanium Grade 5 and one titanium Grade 2, featuring a $ \SI{50}{\mm} $ diameter hemispherical geometry with a $ \SI{15}{\mm} $ long cylinder.
The probe assembly, sketched in Fig.~\ref{fig:figure1b}, included a graphite support, an alumina-silica insulator, and a set of zirconia screws to hold the sample in place.
The low thermal conductivity of both the insulator ($ \sim \SI{0.4}{\watt/\meter\mbox{-}\kelvin} $) and screws ($ \sim \SI{2.2}{\watt/\meter\mbox{-}\kelvin} $) ensured a limited heat transfer to the cooled probe holder.
This probe assembly was preferred with respect to the ESA Standard probe geometry, employing $\diameter \SI{26}{\mm}$ coin-shaped samples, since previous studies revealed pronounced thermal losses between metallic samples and the probe support in the latter case \cite{Fagnani2019}. 

A thermocouple (TC) probe was inserted into a $ \diameter \SI{2.5}{\mm} $ hole, machined along the axis of the sample and extending until 5~mm below the surface. 
A $ \diameter \SI{0.5}{\mm} $, $ \SI{10}{\cm} $ long type-C (W5Re-W26Re) TC wire (Concept Alloys, USA), calibrated until $ \SI{2300}{\degreeCelsius} $, was inserted into a double-bore alumina tube, and a Macor support allowed to house the connection to a dedicated extension wire until the data acquisition unit.
Raw voltage was compensated for the cold-junction value and converted to temperature through the calibration data provided by the manufacturer.

Before testing, the material samples were cleaned in an ultrasonic bath for 30 min with a solution of ammonia, rinsed in demineralized water and finally dried at room temperature.
Within 30~min after testing, samples were stored in a protective argon atmosphere to minimize further oxidation and contamination before the subsequent surface analysis.
Scanning Electron Microscopy (SEM)
images were collected with a JSM IT-300 (Jeol Ltd., JP), characterized
by a lateral resolution of 20 nm at 20 kV, and equipped with an Oxford
Energy-Dispersive X-ray Spectroscopy (EDS) detector.

\subsection{Plasma flow characterization and test conditions}
The characterization of the plasma flow followed the experimental-numerical procedure presented in Ref.~\cite{fagnani2026}, which is summarized here.
An optical emission spectroscopy (OES) set-up (not shown in Fig.~\ref{fig:figure6_3b}) was used to measure spatially-resolved emission spectra of the free-jet plasma flow, allowing the determination of the local gas temperature, $\Ts$, from atomic oxygen lines near 777~nm. The assumption of thermal and chemical equilibrium was supported by complementary measurements, including spectral fits and electron number density obtained from UV to NIR measurements.
The jet axial velocity, $u\subrm{s}$, was inferred from the measured dynamic pressure,  $\pdyn$, accounting for low Raynolds number effects.
The reacting stagnation line flow was then numerically simulated from the experimental input values of $\Ts$, $\us$ and equilibrium flow composition imposed at a distance $\deltastar$ from the probe. Solving the quasi-1D Navier–Stokes equations within this framework, assuming thermal equilibrium and chemical non-equilibrium, yields predicted cold-wall heat fluxes in agreement with the measured values over a selected range of operating conditions, while also enabling estimation of the velocity gradient, $\beta\subrm{w}$, required to identify the corresponding flight conditions \cite{turchi2021}.
Compared to a traditional procedure based on the inverse heat transfer method, the aforementioned methodology is independent of assumptions on the catalytic efficiency of the copper surface and kinetic rates used to model the non-equilibrium gas chemistry within the boundary layer, thus providing a higher confidence in determining the flow conditions summarized in Table~\ref{tab:titanium_conditions}.

Flight conditions, in terms of velocity $ V\subrm{f}$, altitude $ z\subrm{f} $, and radius $ R\subrm{f}$ are obtained from matching the boundary layer edge enthalpy, pressure and velocity gradient \cite{kolesnikov2000} and are reported in Table~\ref{tab:titanium_flight_conditions}.
Test TiG5-D  and TiG2-A were selected to reproduce enthalpy levels and flight radii of a representative $ \SI{50}{\cm} $ diameter tank. 
A milder enthalpy level for test TiG5-B, instead, allowed to observe oxidation in more detail on a longer test duration.
After reaching the desired test conditions in terms
of chamber pressure and electric power, OES of the free-stream flow was performed and the cold-wall heat flux
and jet dynamic pressure were measured. 
Successively, the sample was injected into the plasma flow for a test time $ \Dt $, before switching off the torch. 
The test time was about 66 and 62~s for TiG5-D and TiG2-A, respectively, while a longer test time of 364~s was allowed for TiG5-B.

\begin{table}[]
	\centering
	\caption{Experimental conditions for the tests reported in this work.}
	\resizebox{\textwidth}{!}{  
		\begin{tabular}{ll|llll|lllll|ll}
			\multirow{2}{*}{Test ID} & $ \Delta m $ & $ \mdot $ & $ \Pel $ & $ \pc $ & $ z $ & $ \Ts $ & $ \hs $ & $ \qcw $ & $ \pdyn $ & $ \Delta t $ & $ u\subrm{s} $ & $ \deltastar $ \\
			~ & $ \SI{}{\gram} $ & g/s & kW & mbar & mm & K & MJ/kg & $ \SI{}{\kW/\meter\squared} $ & Pa & s & m/s & mm \\ 
			\midrule
			TiG5-B & 0.34* & 16 & 125 & 50 & 385 & 5260 & 14.31 & 500 & 31 & 364 & 152 & 91 \\ 
			TiG5-D & 0.55 & 16 & 200 & 50 & 385 & 5935 & 24.35 & 1430 & 53 & 66 & 230 & 62 \\ 
			TiG2-A & 0.81 & 16 & 200 & 50 & 385 & 5910 & 23.89 & 1360 & 51 & 62 & 226 & 63 \\ 
			\hline
			\multicolumn{13}{l}{*Part of the oxide crust detached during cool-down and it could not be weighted.}
		\end{tabular}
	}
	\label{tab:titanium_conditions}
\end{table}

\begin{table}[]
	\centering
	\small
	\caption{Scaled flight conditions for the test conditions reported in table \ref{tab:titanium_conditions}.}
	\resizebox{0.6\textwidth}{!}{  
		\begin{tabular}{llllll}
			\multirow{2}{*}{Test ID} & $ \hs $ & $ \beta\subrm{w} $ & $ V\subrm{f} $ & $ z\subrm{f} $ & $ R\subrm{f} $ \\ 
			~  & MJ/kg & 1/ms & km/s & km & cm  \\ 
			\midrule
			TiG5-B  & 14.31 & 5.89 & 5.34 & 64.5 & 34.2  \\ 
			TiG5-D  & 24.35 & 9.32 & 6.98 & 68.4 & 25.1  \\ 
			TiG2-A  & 23.89 & 9.11 & 6.90 & 68.3 & 25.6 \\ 
			\hline
		\end{tabular}
	}
	\label{tab:titanium_flight_conditions}
\end{table}

\subsection{In situ multi-band emissivity measurements}
Time-resolved multi-band angular emissivity measurements during exposure to the plasma flow are based on the method described in Ref.~\cite{fagnani2024}.
This leverages a general model for the response of a bandpass infrared
radiometer, including detector spectral sensitivity, optics, and atmospheric path transmission.  
We used the KT19 broadband radiometer (Heitronics Infrarot Messtechnik Gmbh, DE) in the  $ \IRMband{0.6}{39} $ range, the FLIR A6750sc (Teledyne FLIR LCC, USA) infrared camera in the $ \IRMband{3}{5.3} $ range, the Optris CTG5 (Optris GmbH, DE) within the $ \IRMband{4.8}{5.2} $ range, and the Marathon Series MR1SB (Raytek Corporation, USA) two-color pyrometer in the near-infrared wavelength range, between $ 0.75\textit{-}\SI{1.1}{\micro\meter} $ and $ 0.95\textit{-}\SI{1.1}{\micro\meter} $. 
The material surface temperature was simultaneously measured from the ratio of the latter two signals, according to the procedure described in Ref.~\cite{fagnani2024a}, and referred to as Two-Color Pyrometry (TCP).
A dedicated analysis revealed necessary to corroborate such temperature measurements for the oxidizing metallic surface, and it is discussed in \ref{sec:appendix_TCP}.

The radiometric calibration of the instruments was performed with a
variable temperature blackbody source (Landcal R1500T, Ametek Land, UK) in the range 773-1773~K, and extrapolated to 2200~K.
Table~\ref{tab:instrument_parameters} summarizes the main parameters of the instrument set-up.
The instruments were positioned at a distance of $ \SI{1}{\meter} $ from the test sample, with an inclination of $ 50^{\circ} $ (view-ports A, B and C) and  $ 35^{\circ} $ (view-port D) with respect to the sample surface normal, respectively.
A LabVIEW interface allowed synchronized recording of the radiometers at a frequency of $ \SI{10}{\hertz} $, whereas the IR camera frames were recorded at $ \SI{2}{\hertz} $ and synchronized to the other instruments using a signal generator.

It is important to note that, as discussed in Ref.~\cite{fagnani2024}, the measured emissivity depends on the specific spectral response of the detector. For this reason, the reported quantity is referred to as the apparent directional band emissivity, $\epsstar$, which may differ from the true band emissivity of the material. For the detectors employed in the present work, however, this deviation remained limited to only a few percent, even for metallic surfaces exhibiting a strong wavelength dependence in emissivity \cite{fagnani2024}. Additionally, due to the finite size of the radiometer measurement area, the reported values of surface temperature and emissivity correspond to the average radiance collected over the probed surface spot ($ \diameter 10\textit{--}\SI{17}{\mm} $).

\begin{table*}[]
	\small
	\centering
	\caption{Summary of the instruments and optical path parameters.}
	\resizebox{\textwidth}{!}{
		\begin{tabular}{llllllllllll}
			\hline
			instrument & detector & $ \Delta \lambda $ & $ \Delta t $  & spot size & window & label & $ \taustarext $ & $ d_1 $ & $ d_2  $& $ d $ & $ \alpha $ \\ 
			~ & ~ & $ \SI{}{\um} $ & $ \SI{}{\ms} $ & $ \SI{}{\mm} $ & ~ & ~ & ~ & $ \SI{}{\cm} $ & $ \SI{}{\cm} $ & $ \SI{}{\cm} $ & deg \\ \hline
			KT19 & pyroelectric &  0.6-39 & 100 & 14  & KRS5 & B & 0.70 & 90 & 10 & 100 & 50 \\
			\multirow{2}{*}{MR1SB} & \multirow{2}{*}{$\ce{Si}/\ce{Si}$ layered} & 0.75-1.1 & \multirow{2}{*}{10} &  \multirow{2}{*}{14} & \multirow{2}{*}{quartz} & \multirow{2}{*}{D} & \multirow{2}{*}{0.87} &\multirow{2}{*}{80} & \multirow{2}{*}{20} & \multirow{2}{*}{100} & \multirow{2}{*}{35} \\
			& & 0.95-1.1 & &  &  & & &  &  \\ 
			A6750sc & $ \ce{InSb} $ & 3-5.3 & 1 & 10 & $ \ce{CaF2} $ & A & 0.95 & 90 & 10 & 100 & 50 \\
			CTG5 & thermopile & 4.8-5.2 & 80 & 17 & $ \ce{CaF2} $ & C & 0.95 & 90 & 10 & 100 &  50 \\ 
			\hline
		\end{tabular}
	}
	\label{tab:instrument_parameters}   
\end{table*}

\subsection{Spatially calibrated high-resolution camera frames}

We used the Prosilica GT2050C (Allied Vision, USA) featuring a 4 megapixel sensor and coupled to a 200~mm objective to image the samples during exposure to the plasma flow at a frequency of 4~Hz. 
The spatial calibration approach developed in Ref.~\cite{fagnani2023a} for infrared thermography was extended here to high-resolution images in the visible range. Camera calibration relied on a pinhole model with lens distortion correction implemented in OpenCV \cite{opencv_library}, using a dedicated calibration target with known control points. This procedure enabled the mapping of 2D image data onto the three-dimensional sample geometry, thereby allowing quantification of the spatial evolution of the oxide layer formed during plasma exposure.

%
\section{Results and discussion}
\graphicspath{{figures/}}

\subsection{Post-test visual inspection and microscale analysis}

\begin{figure}[]
	\includegraphics[trim={0cm, 0cm, 1cm, 0cm}, clip, width=0.95\textwidth]{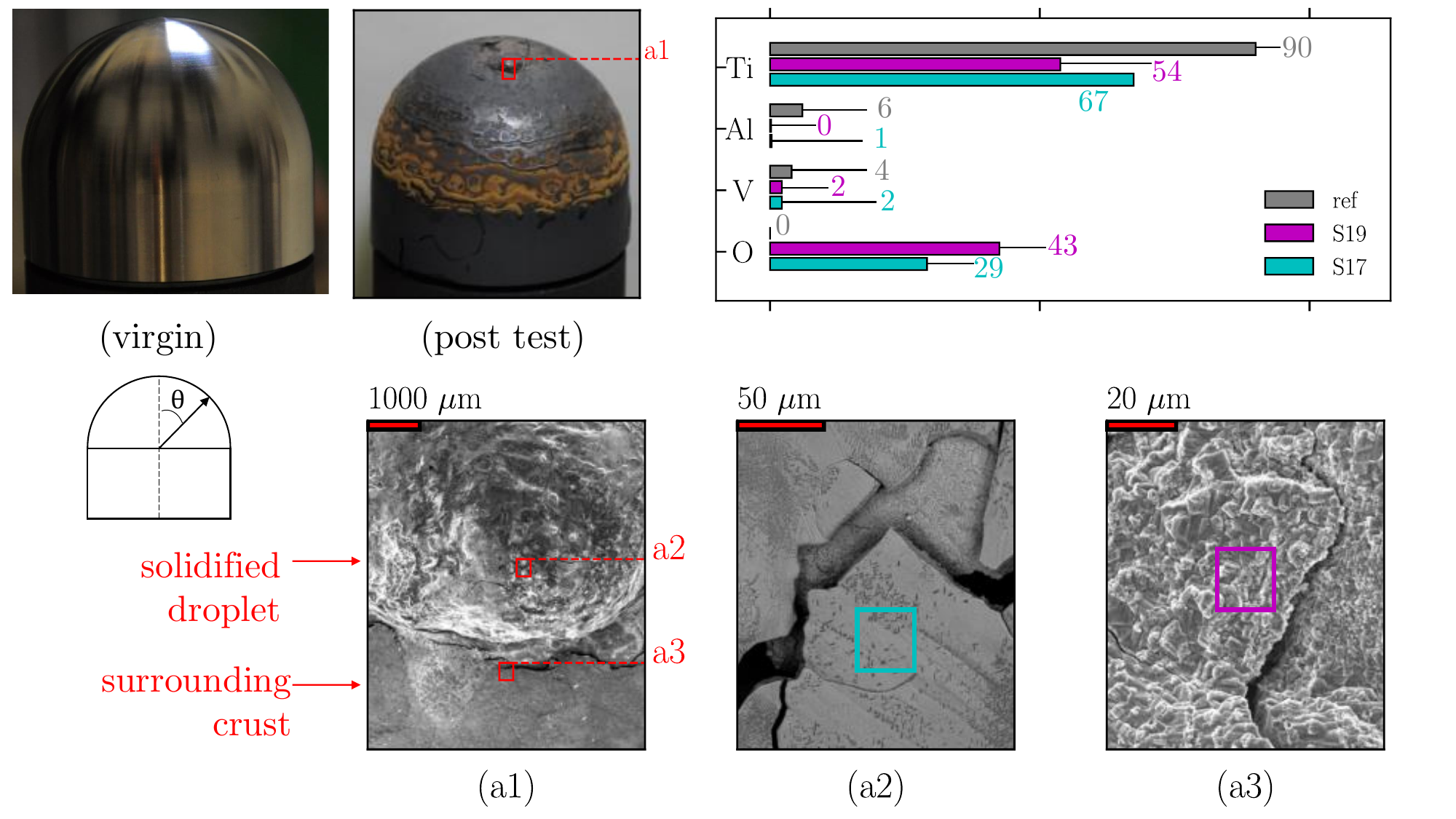}
	\caption{TiG5-D: virgin and post-test pictures of the sample. SEM micrographs: area around the solidified droplet and surrounding crust (a1); detail of the solidified droplet (a2); detail of the surrounding crust (a3). Bar chart reports composition of salient elements in wt\% from EDS of areas indicated with corresponding colors (grey bars are reference Ti-6Al-4V composition).}
	\label{fig:figure2} 
\end{figure}

\begin{figure}[]
	\includegraphics[trim={0cm, 0cm, 1.5cm, 0cm}, clip, width=0.95\textwidth]{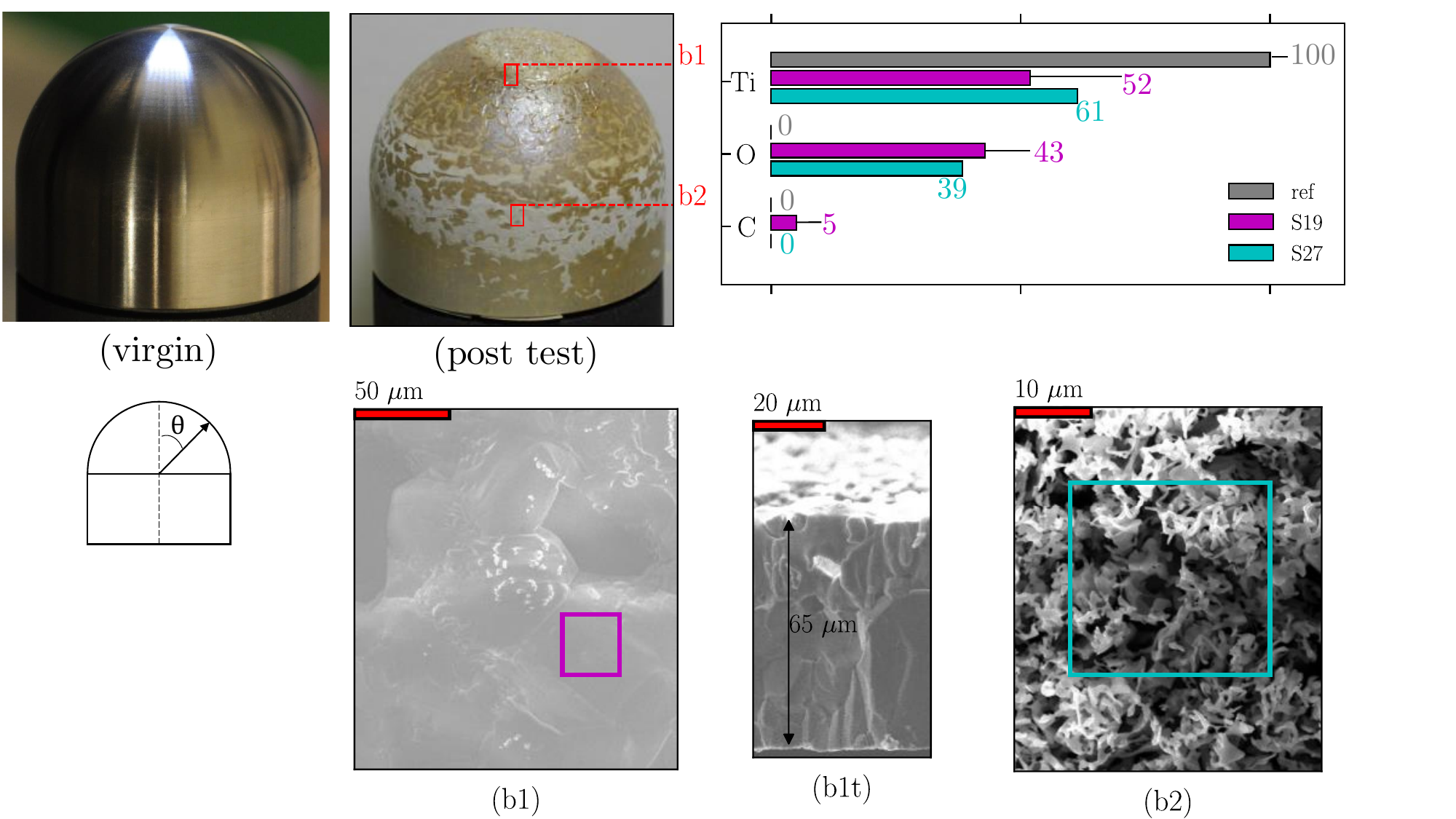}
	\caption{TiG2-A: virgin and post-test pictures of the sample. SEM micrographs: area around the solidified region (b1) shows compact smooth crust of about $ \SI{65}{\um} $ thickness (b1t); white area around the hemisphere-cylinder edge shows porous filament structure (b2).  
		Bar chart reports composition of salient elements in wt\% from EDS of areas indicated with corresponding colors (grey bar is reference Ti composition).}
	\label{fig:figure3}
\end{figure}

\begin{figure}[h!]
	\centering
	{\includegraphics[trim = {0cm, 12cm, 0.5cm, 0.cm}, clip, width=0.95\linewidth]{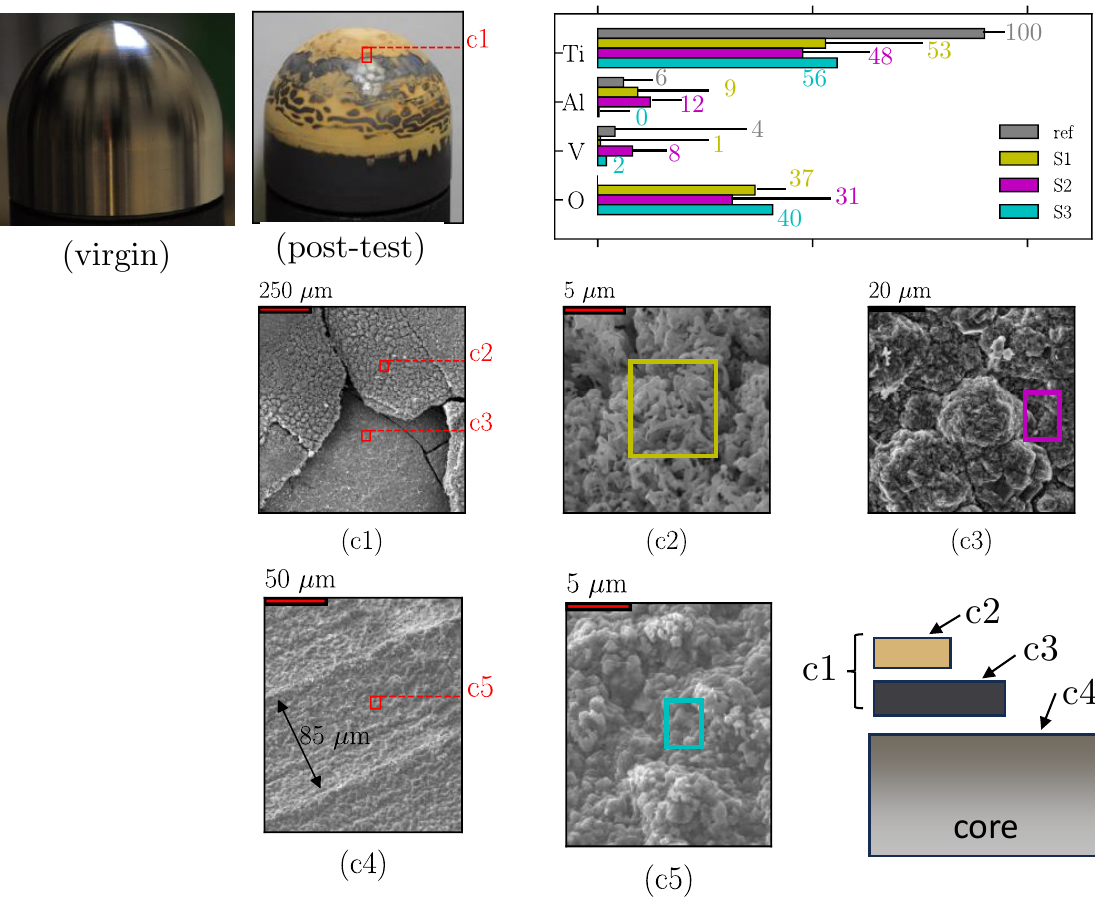}}
	\caption[Pre- and post-test picture, SEM and EDS of TiG5-B.]{TiG5-B: virgin and post-test pictures of the sample. SEM micrographs: the crust shows a duplex layer (c1); yellow crust shows filament structure (c2); underlying brown layer is compact with visible grains (c3); surface of the core material showing machining grooves about $ \SI{85}{\um} $ wide (c4); detail of the surface with visible oxide crystals (c5). Bar chart reports composition of salient elements in wt\% from EDS of areas indicated with corresponding colors (grey bars are reference Ti-6Al-4V composition).}
	\label{fig:figure4}
\end{figure}

Figures~\ref{fig:figure2},~\ref{fig:figure3},~and~\ref{fig:figure4} compare the virgin and post-test appearance of the samples, revealing diverse oxidation morphology with distinct surface patterns across different regions. 
Unlike previous studies performed in dissociated oxygen atmospheres~\cite{Prevereaud2016, Balat-Pichelin2020}, no significant oxide delamination was observed during plasma exposure. Although the oxide crust appeared brittle, it remained largely intact for both TiG5-D and TiG2-A, enabling reliable mass-gain measurements of 0.55~g and 0.81~g, respectively. In contrast, sample TiG5-B experienced partial crust detachment during the cooling phase, which also fractured easily during post-test handling, resulting in a non-representative mass gain of 0.34~g.

For TiG5-D (Fig.~\ref{fig:figure2}), a black crust extends up to $ \theta \approx \SI{60}{\degree} $, transitioning to a yellow region for $ 60 < \theta < \SI{90}{\degree} $. SEM micrographs show the boundary between the molten and oxidized zones (a1): the molten surface appears compact with irregular cracking (a2), while the surrounding crust forms a thin, partially detached layer (a3). EDS analysis confirms extensive oxidation, with Ti/O ratios of 2.3 in the molten area and 1.26 in the crust, suggesting the coexistence of TiO and $\ce{TiO2}$ phases. Minor amounts of Al and approximately 2\% V were also detected.

Sample TiG2-A (Fig.~\ref{fig:figure3}) exhibits a uniform golden crust interspersed with white spots. The crust is dense in the melted region (not analyzed by SEM/EDS) and compact in adjacent zones (b1), with a thickness of approximately $\SI{65}{\micro\meter}$ (b1t). The white regions display a porous network of filaments 5--$\SI{10}{\micro\meter}$ long (b2). EDS results indicate pronounced oxidation and carbon enrichment in the compact areas.

The TiG5-B sample (Fig.~\ref{fig:figure4}) presents a yellow crust with black speckled features. A section between 30 and $\SI{45}{\degree}$ along the hemispherical region detached during cooling, exposing the substrate. SEM analysis at the fractured edge reveals multiple oxide layers (c1). The upper yellow layer (c2) displays a porous, filamentary structure similar to that observed in (b2), overlaying a compact black layer with granular features 5--$\SI{30}{\micro\meter}$ in size (c3). The presence of such duplex oxide scales is consistent with previous observations reported in the literature \cite{Wei2012, Balat-Pichelin2020, brault2025}. Macroscopically, the two oxide layers remain adherent to each other but are mostly separated from the underlying substrate (c4), which retains grooves about $\SI{85}{\micro\meter}$ wide, likely originating from machining process, and crystal grains near $\SI{1}{\micro\meter}$ (c5). EDS data reveal oxygen contents between 30--40\%, with Al and V enrichment in the detached layers: Al and V are both abundant in the compact oxide (c3), while the yellow porous layer (c2) shows V depletion and Al enrichment.

Overall, porous morphologies such as those in (b2) and (c2) likely enhance subsurface oxidation through increased oxygen diffusion. Nevertheless, the Al and V concentrations varied significantly even among neighboring regions. Importantly, no nitrogen was detected on the analyzed surface, indicating negligible Ti--N reactivity under the tested conditions, consistent with the limited atomic nitrogen fraction expected in the boundary layer for edge temperatures below 6000~K.

\newpage
\subsection{In situ material response analysis}

\begin{figure}[h!]
	\centering
	\includegraphics[width=0.99\linewidth, trim={1.7cm, 1.3cm, 0.5cm, 0cm}, clip]{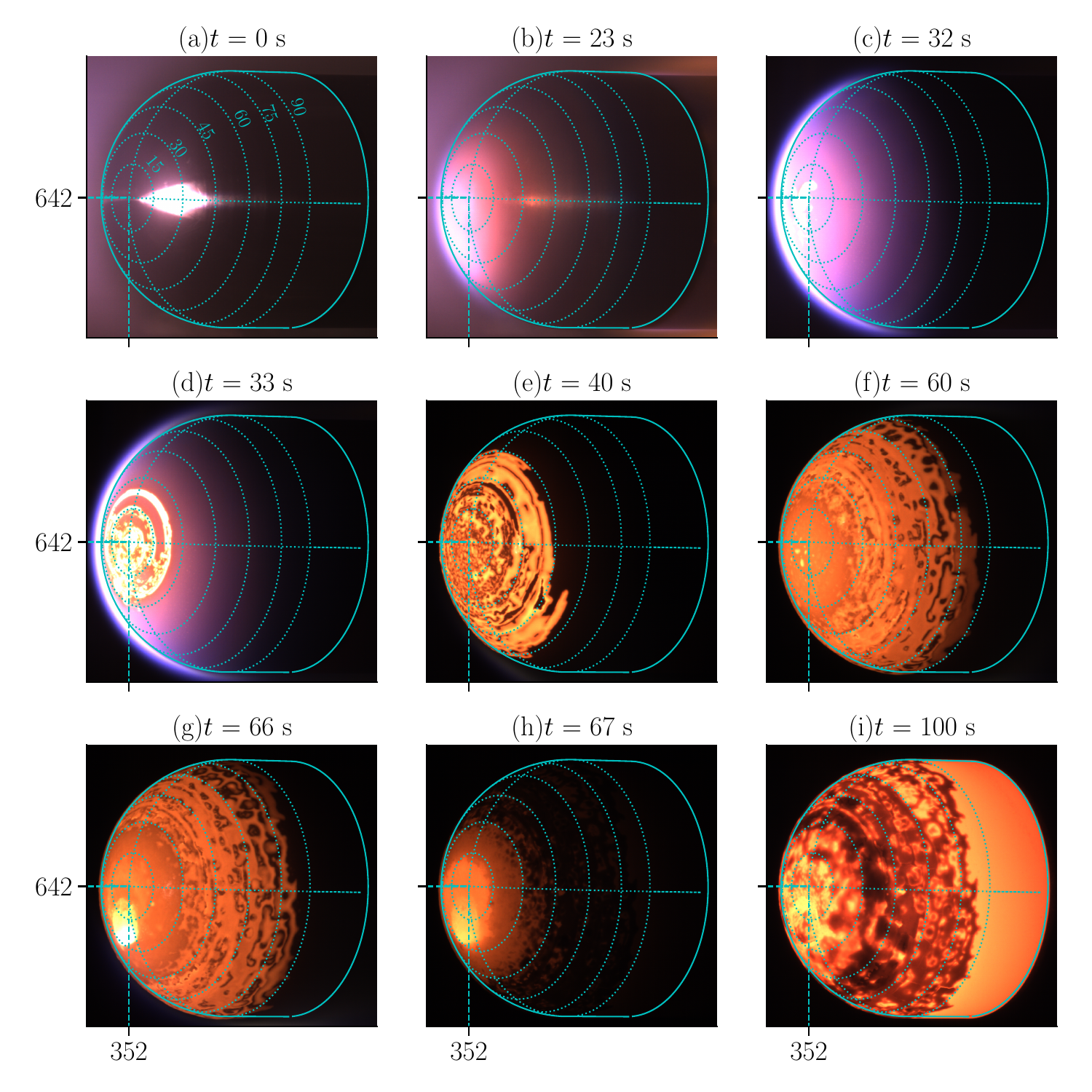}
	\caption[]{Spatially calibrated camera frames during test TiG5-D, corresponding to the time history in Fig.~\ref{fig:figure6a}. Cyan lines highlight the sample geometry and the angular position over the hemispherical part (indicated in the first frame). After injection (a), temperature reading starts at 23~s (b). The sudden formation of a thick oxide crust at 32~s (c-d) corresponds to the observed temperature jump. The crust evolves to $ \theta \sim \SI{90}{\degree} $ at $ t = \SI{60}{\second} $ (d-f), when a liquid droplet forms beneath (f-g). Cool down after $ t = \SI{67}{\second} $ (h-i) highlights different contrast patterns of the crust.}
	\label{fig:figure5}
\end{figure}

\begin{figure}[h!]
	\centering
	\subfigure[TiG5-D]
	{\includegraphics[width=0.9\linewidth]{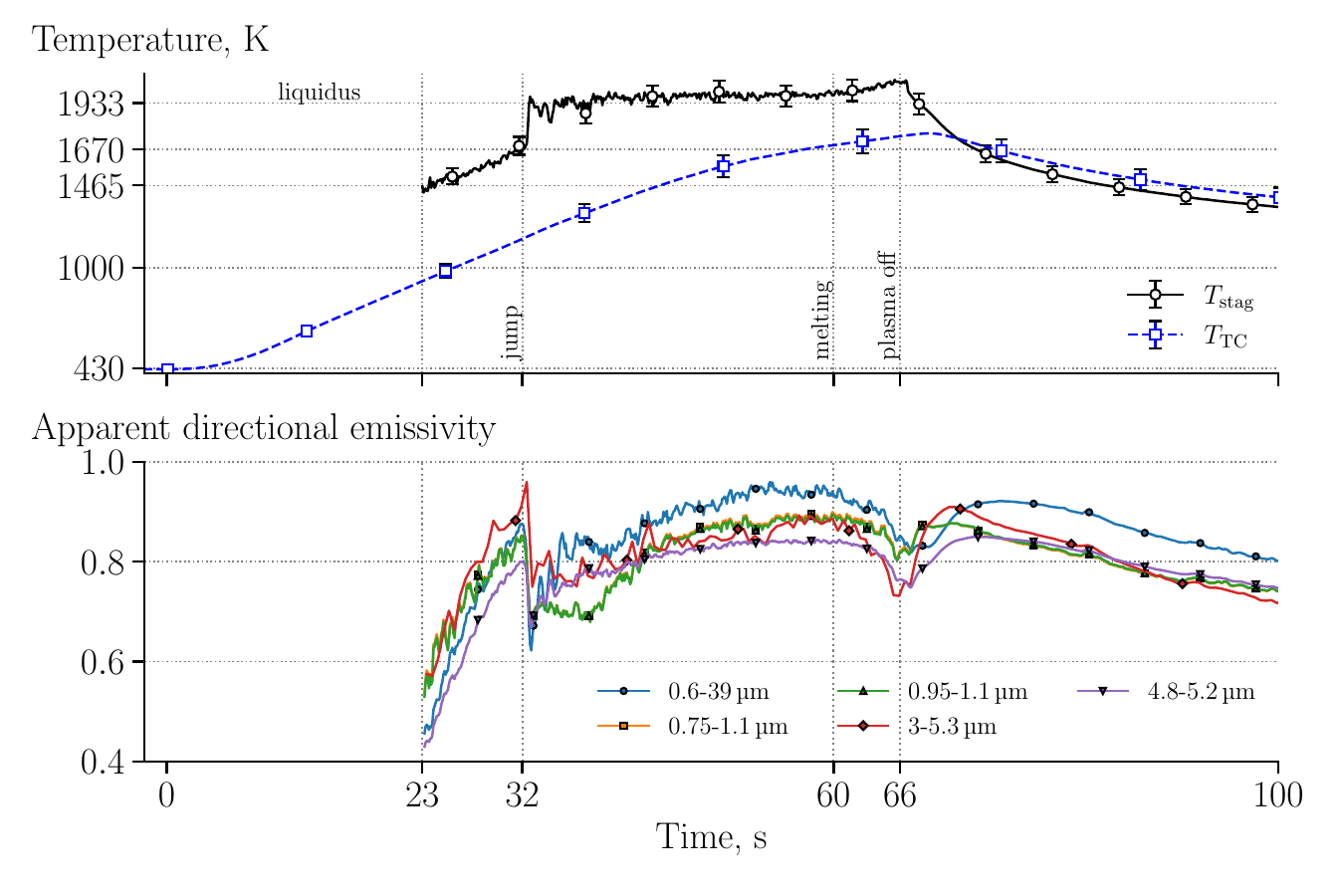}
		\label{fig:figure6a}}\\
	\subfigure[TiG2-A]
	{\includegraphics[width=0.9\linewidth]{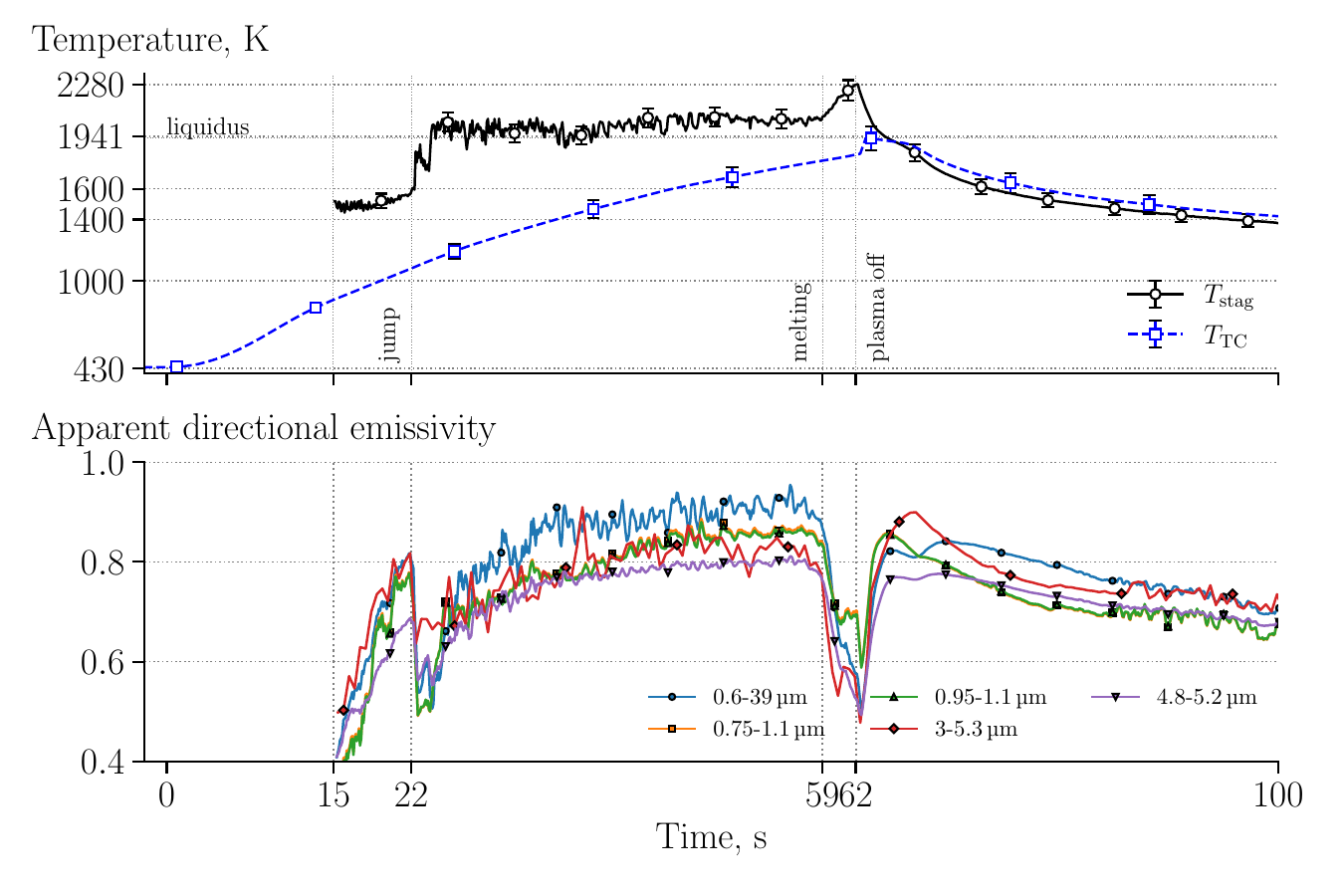}\label{fig:figure6b}}
	\caption[]{Evolution of the surface temperature at stagnation region ($ \Tstag $), at 5~mm below the surface ($ \Ttc $), and measured in situ apparent angular emissivity, $ \epsstar $, within different bands for samples TiG5-D (a) and TiG2-A (b). $ \epsstar $ increases until a sudden jump in temperature occurs, causing it to decrease. $ \Tstag $ rises at the onset of melting, while $ \epsstar $ drops, to recover larger values after solidification during the cool-down phase.}
	\label{fig:TiG5-G2_time_T_eps}
\end{figure}

For test TiG5-D, Fig.~\ref{fig:figure5} presents spatially calibrated camera frames at selected instants, while Fig.~\ref{fig:figure6a} illustrates the temporal evolution of the stagnation-point surface temperature ($ \Tstag $), the thermocouple temperature measured 5~mm below the surface ($ \Ttc $), and the corresponding apparent angular band emissivity ($ \epsstar $) across different radiometer bands. 
Uncertainty values on $ \epsstar $ are omitted from the plot for clarity and quantitative values as a function of temperature are provided in Ref.~\cite{fagnani2024}.

At injection, $ \Ttc $ indicates an initial temperature near 430~K. Reliable surface temperature measurements from TCP became available only after approximately $\SI{23}{\second}$, when $ \Tstag $ reached $\SI{1465}{\kelvin}$. 
The surface temperature then increased steadily to $\SI{1670}{\kelvin}$, followed by a sudden jump to $\SI{1930}{\kelvin}$ -- close to the liquidus temperature of the alloy. 
During this phase, $ \epsstar $ gradually increased from $\sim0.5$ to $\sim0.8$ across all spectral bands, before dropping abruptly at the temperature jump. 
Simultaneously, camera frames in Fig.~\ref{fig:figure5}(c--d) reveal the emergence of a bright spot at the stagnation region, expanding rapidly to $\theta \sim \SI{30}{\degree}$ within $\SI{1}{\second}$. 
The prior increase in $ \epsstar $ suggests ongoing oxidation, while the sudden emissivity drop corresponds to the formation of a thick, bright oxide crust exhibiting reduced emissivity.

As $ \Tstag $ stabilizes near $\SI{1980}{\kelvin}$, $ \epsstar $ increases again to values between 0.8 and 0.9. The oxide front extends progressively, reaching $\SI{45}{\degree}$ at $\SI{40}{\second}$ and $\SI{90}{\degree}$ at $\SI{60}{\second}$. 
At this stage, a molten layer develops beneath the oxide scale. The surface temperature further rises to $\SI{2050}{\kelvin}$, accompanied by a simultaneous decrease in emissivity across all radiometer bands. This behavior is consistent with a lower emissivity of the molten phase. 
Upon plasma shut-off at $\SI{66}{\second}$, the material resolidifies and $ \epsstar $ partially recovers, gradually declining to about 0.6 during cooldown. 
Notably, the subsurface temperature $ \Ttc $ remains unaffected by the abrupt $ \Tstag $ transition, increasing smoothly to $\sim \SI{1800}{\kelvin}$ when the plasma is switched off.

A comparable thermal and radiative behavior is observed for the Grade~2 titanium sample TiG2-A, shown in Fig.~\ref{fig:figure6b}. 
Both $ \Tstag $ and $ \epsstar $ exhibit similar trends, and the oxide crust evolved in an analogous manner (not shown), confirming that the observed temperature-emissivity jump is not specific to Grade~5 but likely results from the intrinsic Ti--O interaction under these conditions. 
The main difference lies in the timing and intensity of the transition: for TiG2-A, the jump occurs earlier (at $\SI{22}{\second}$) and at a slightly lower temperature ($\SI{1600}{\kelvin}$), about 10~s sooner and 70~K below that of TiG5-D. 
Moreover, the emissivity drop during melting (between 59 and 62~s) is more pronounced, with $ \Tstag $ peaking near $\SI{2280}{\kelvin}$. 
It should be noted that, due to the spatial non-uniformity of the oxidized surface, the reported values of $ \Tstag $ and $ \epsstar $ represent integral averages of the radiance collected over the measurement spot ($ \diameter 10\textit{--}\SI{17}{\mm} $). 

\begin{figure}[]
	\centering
	{\includegraphics[width=0.99\linewidth]{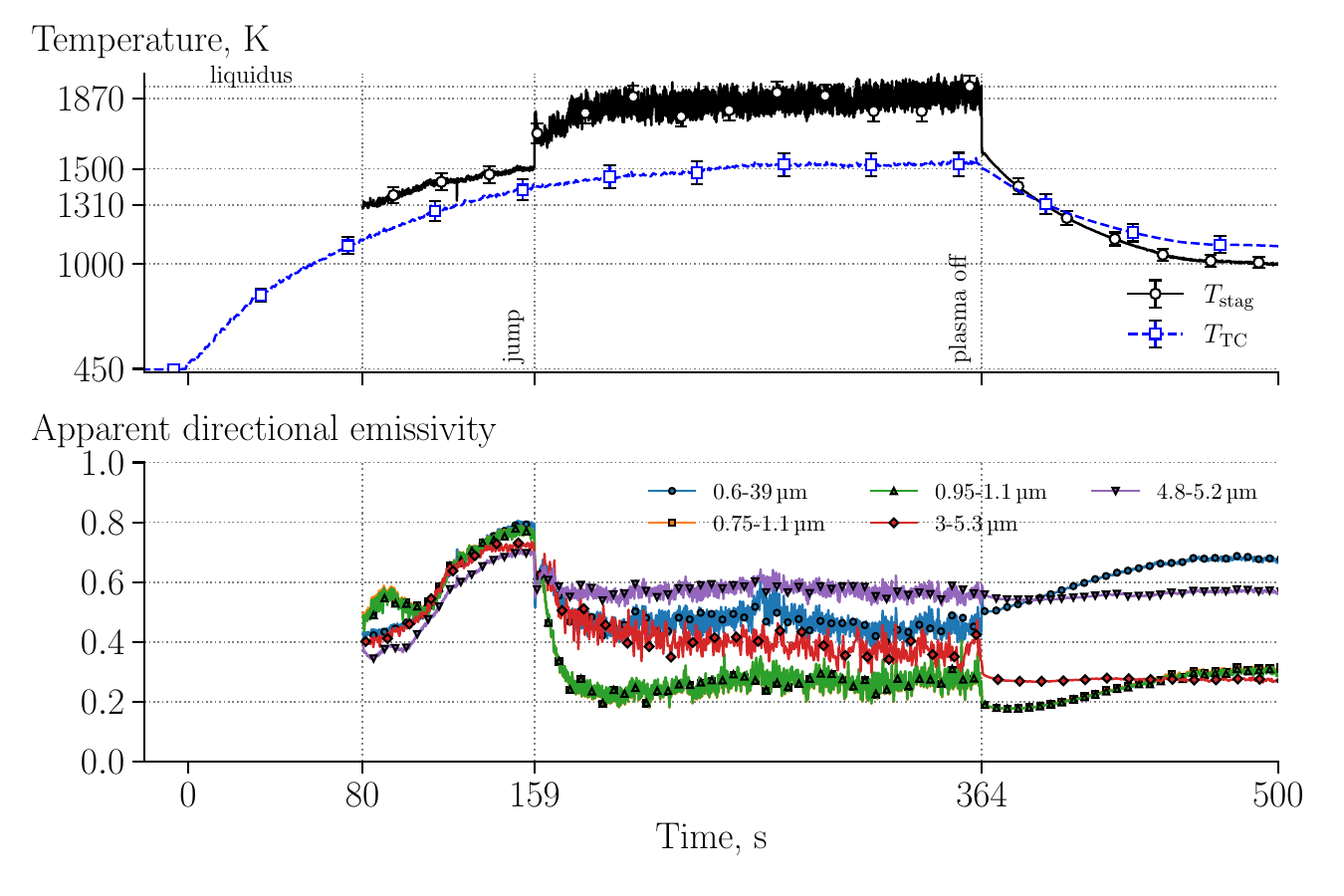}}
	\caption[]{TiG5-B: Evolution of the surface temperature at stagnation region ($ \Tstag $), at 5~mm below the surface ($ \Ttc $), and measured in situ apparent angular emissivity, $ \epsstar $, within different bands. $ \epsstar $ suddenly decrease as temperature jumps, settling to considerably lower values than those in Fig.~\ref{fig:figure6a}, which are almost constant during cool down. }
	\label{fig:figure7}
\end{figure}

A milder test condition, featuring a free-stream enthalpy of 14.28~MJ/kg and a longer exposure time of approximately 364~s, enabled a clearer observation of the temperature-jump phenomenon on sample TiG5-B. 
Fig.~\ref{fig:figure7} presents the recorded stagnation-point surface temperature ($ \Tstag $), subsurface thermocouple temperature ($ \Ttc $), and the corresponding apparent angular band emissivity ($ \epsstar $). 
Reliable TCP surface temperature readings were obtained only after $\SI{80}{\second}$, when $ \Tstag $ reached $\SI{1310}{\kelvin}$. 
The temperature increased gradually to $\SI{1500}{\kelvin}$ after $\SI{159}{\second}$, followed by a sudden jump to $\SI{1680}{\kelvin}$, and then stabilized around $\SI{1870}{\kelvin}$ until plasma shut-off. 
As in previous tests, the subsurface thermocouple signal -- measured 5~mm below the surface -- showed no abrupt slope changes, rising steadily to slightly above $\SI{1500}{\kelvin}$. 
The lower jump temperature observed here compared to TiG5-D suggests that the phenomenon is not purely temperature-driven, but rather linked to the flow and/or specific oxidation conditions.

Similar to TiG5-D, the emissivity $ \epsstar $ increased from approximately 0.4 to 0.8 across all spectral bands prior to the temperature jump, after which it decreased sharply, exhibiting distinct wavelength dependence. 
Within roughly 30~s after the jump, $ \epsstar $ reached a quasi-steady plateau that persisted until the end of plasma exposure. 
Unlike TiG5-D, however, emissivity values remained significantly lower overall, particularly in the $\IRMband{0.75}{1.15}$ band, where $ \epsstar $ stabilized near 0.3.

\begin{figure}[h]
	\centering
	\includegraphics[width=0.99\linewidth, trim={1.7cm, 1.3cm, 0.5cm, 0cm}, clip]{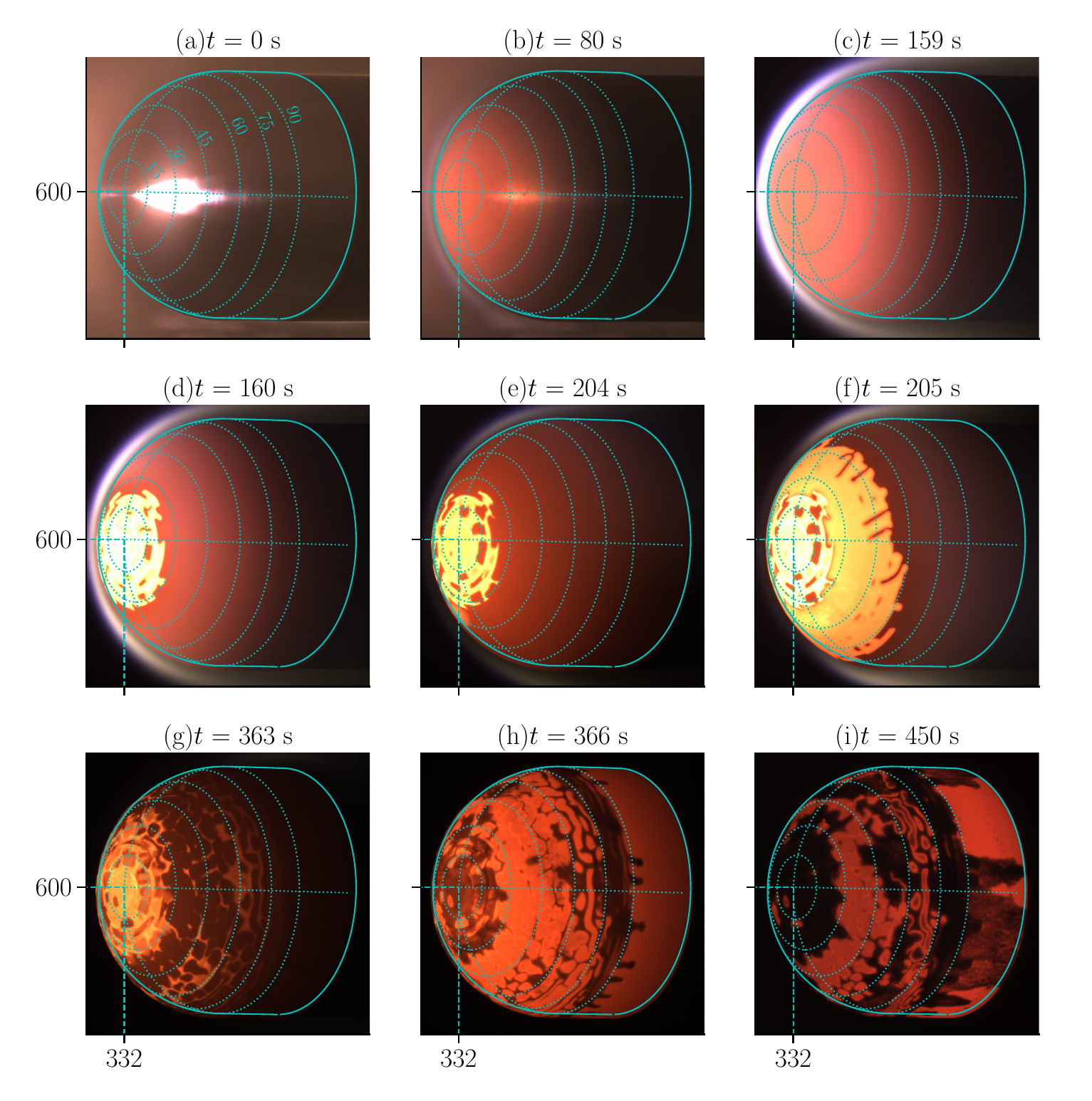}
	\caption[Image sequence showing oxidation of TiG5-B.]{Image sequence during test TiG5-B: (a) injection, (b) temperature reading starts and regular heating until 159~s (c). A bright spot appears as temperature jumps at 160~s (d), attributed to the appearance of the yellow crust visible in Fig~\ref{fig:figure4}. The crust evolves abruptly between 204 and 205~s (e-f). Brighter patterns during plasma exposure appear darker during cool-down (h) and possibly indicate detachment and fast cooling of the crust (i).}
	\label{fig:figure9_7}
\end{figure}

The image sequence in Fig.~\ref{fig:figure9_7} confirms the dynamic nature of this oxidation process. 
A bright, irregular pattern extending to $\theta \sim \SI{30}{\degree}$ appeared abruptly between 159 and 160~s and remained largely unchanged until $\SI{204}{\second}$, before expanding rapidly to $\theta \sim \SI{60}{\degree}$ at $\SI{205}{\second}$. 
Near the end of plasma exposure [Fig.~\ref{fig:figure9_7}(g)], the oxidized surface exhibits a speckled texture over the hemispherical region of the sample. 
Upon plasma shut-off [Fig.~\ref{fig:figure9_7}(g–h)], the brightness pattern evolves almost identically to its earlier shape, indicating a pronounced thermal response consistent with rapid cooling of a thin, partially detached surface layer. 
During the cooldown phase, dark areas emerge suddenly, supporting the hypothesis that the oxide crust progressively delaminates from the substrate, likely driven by thermal expansion mismatches and transient stress gradients.

\subsection{Interpretation of the oxidation mechanism}

Combining the temporal evolution of $ \Tstag $, $ \Ttc $, and $ \epsstar $ with the camera observations, post-test inspection, and microscopic analyses, it is evident that both Grade~2 and Grade~5 titanium alloys undergo oxidation through distinct mechanisms under simulated atmospheric-entry conditions.

For all samples, the progressive rise in $ \epsstar $ prior to the temperature jump confirms that oxidation initiates during the heating phase. Compared to available literature showing higher emissivity values of Ti-6Al-4V at lower temperatures \cite{Balat-Pichelin2020a}, a correlation of the heating rate with the onset of oxidation \cite{rajabi2020} can contribute to the delay observed in the present work.
A comparative examination of TiG5-D and TiG5-B indicates that the temperature-emissivity jump and the associated bright pattern observed in the image sequences correspond to the abrupt formation of a yellow surface crust, as shown in Fig.~\ref{fig:figure2} and Fig.~\ref{fig:figure3}. 
This crust subsequently transforms into a black oxide layer as the temperature approaches the melting point ($ \sim \SI{1930}{\kelvin} $), whereas the yellow coloration is maintained only up to mean surface temperatures of approximately $\SI{1870}{\kelvin}$.
The crust forms abruptly at $\SI{1500}{\kelvin}$ and $\SI{1670}{\kelvin}$ for plasma free-jet temperatures of $\SI{5260}{\kelvin}$ and $\SI{5935}{\kelvin}$, respectively, suggesting a dependence on local oxygen concentration and diffusion kinetics within the boundary layer, as well as on the sample heating rate. 
As illustrated in the spatially calibrated frames of Fig.~\ref{fig:figure9_7}(e--f), the crust propagates rapidly over large surface areas once initiated. 
SEM micrographs (Fig.~\ref{fig:figure4}) reveal a highly porous, filamentous morphology superimposed on a compact oxide sublayer, forming a duplex structure with pronounced variations in the relative concentrations of Al and V. 
This likely reflects the onset of a specific thermochemical regime at the surface, where rapid oxidation kinetics promote the growth of filamentary titanium oxides.
Similar multilayer oxide scales where also observed by \citet{Balat-Pichelin2020} and \citet{brault2025}. 

Additionally, the oxide crust appeared solid and adhesive in regions where the surface temperature reached the melting point, whereas it was brittle and prone to detachment elsewhere. The pronounced fluctuations in $ \Tstag $, combined with the absence of a corresponding gradient in $ \Ttc $, particularly for test TiG5-B (see Fig.~\ref{fig:figure7}), indicate that the newly formed oxide likely detaches from the substrate even during plasma exposure before melting occurs. 
The speckled contrast pattern observed in the camera frames further supports this interpretation, showing progressive delamination of the crust during the cooling phase—most likely driven by thermal expansion mismatch between the oxide layer and the underlying metallic substrate.

\subsection{Temperature behavior of emissivity}

\begin{figure}[]
	\centering
	{\includegraphics[trim={0cm, 5.2cm, 0cm, 0cm}, clip, width=0.99\linewidth]{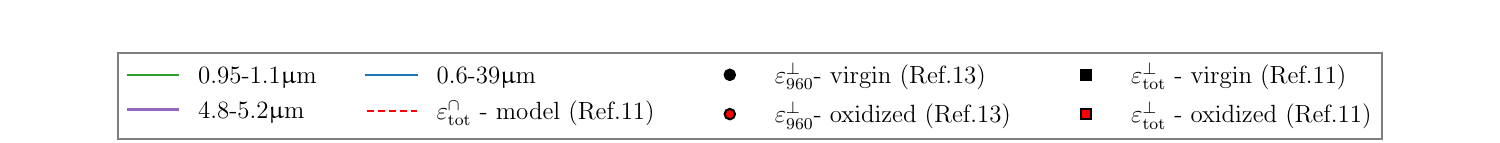}}\\
	\subfigure[TiG5-D: reading starts (A), jump (B-C), settling (C-D), melting (D-E), cool-down (E-F).]
	{\includegraphics[trim={0cm, 6cm, 0cm, 0cm}, clip, width=0.99\linewidth]{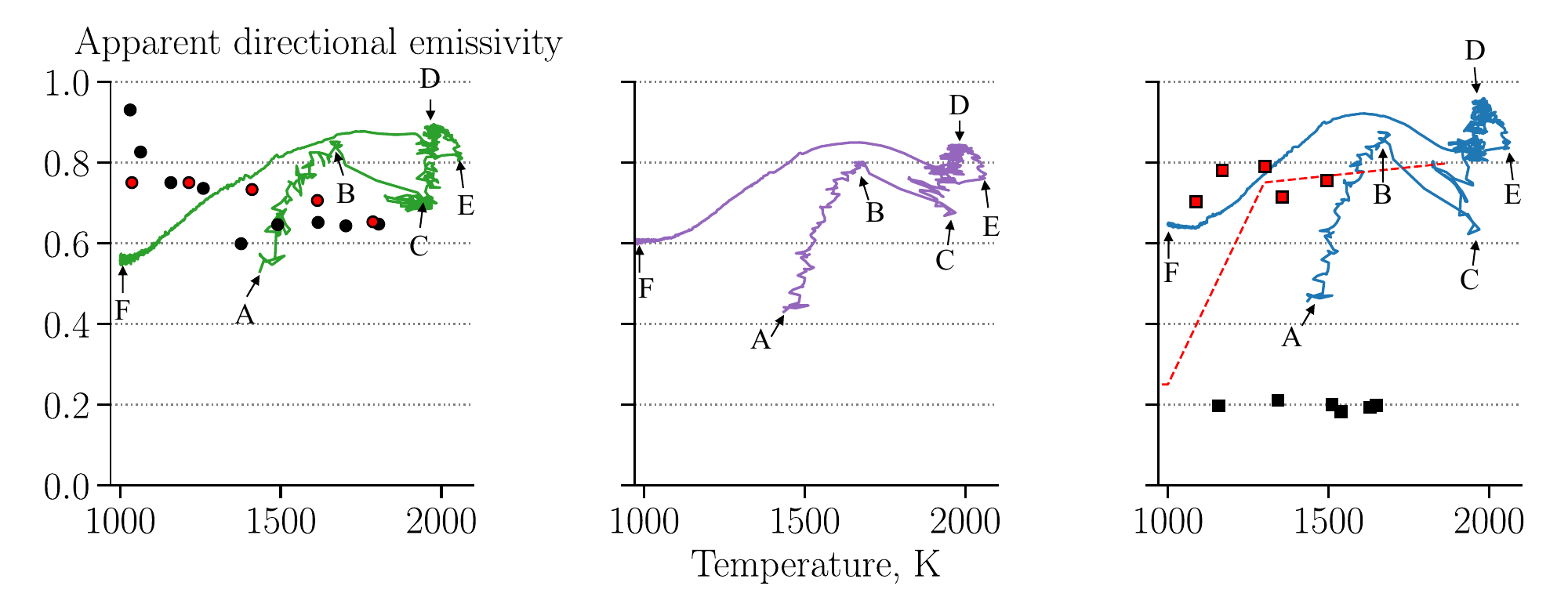}
		\label{fig:figure9_8a}}
	\subfigure[TiG5-B: reading starts (A), jump (B-C), settling (C-D), steady-state fluctuations (D-E), cool-down (E-F).]
	{\includegraphics[trim={0cm, 5cm, 0cm, 0cm}, clip, width=0.99\linewidth]{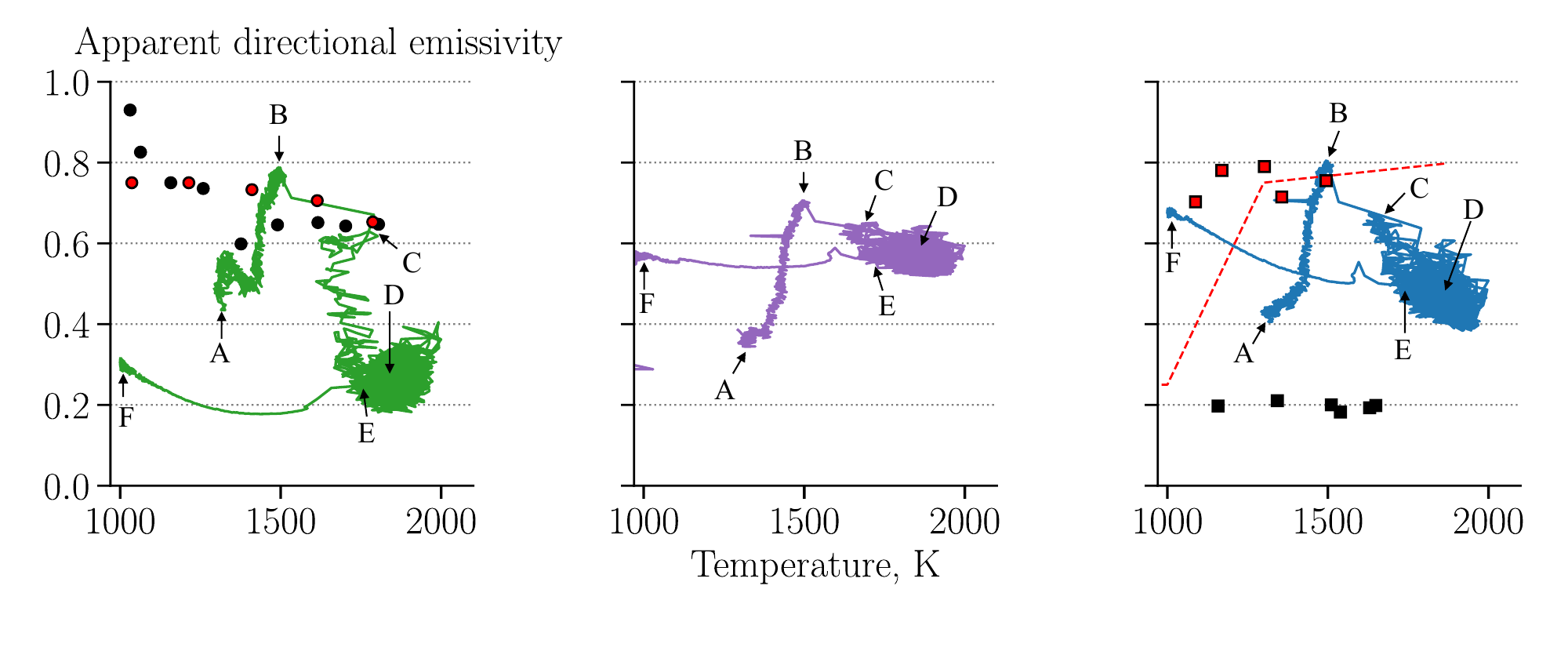}
		\label{fig:figure9_8b}}
	\caption[Temperature evolution of the band emissivity for grade 5 titanium.]{Evolution of $ \epsstar $ with $ \Tstag $ for some selected bands, highlighting the different test phases for TiG5-D (a) and TiG5-B (b). While a similar trend is observed before the jump (A-B), a diverse behavior follows. Present data are additionally compared to values from Refs.~\cite{pagan2016, Balat-Pichelin2020a}. }
\end{figure}

Selected emissivity measurements are presented as a function of the surface temperature for Grade~5 samples in Fig.~\ref{fig:figure9_8a} and Fig.~\ref{fig:figure9_8b}. 
Before (A–B) and across (B–C) the temperature jump, the measured values display similar behavior, though shifted to higher temperatures for TiG5-D. 
After the jump, distinct trends emerge: steady-state emissivity levels for TiG5-B remain substantially lower than those for TiG5-D, reflecting the different evolution of the oxide crust near the melting point achieved in the latter case. 
In particular, the $ \IRMband{0.95}{1.1} $ band [Fig.~\ref{fig:figure9_8b}] shows a pronounced drop -- from $\sim 0.8$ at $\SI{1500}{\kelvin}$ to 0.3 at $\SI{1870}{\kelvin}$ -- whereas the $ \IRMband{4.8}{5.2} $ and $ \IRMband{0.9}{39} $ bands decrease only slightly. 
The observed fluctuations correspond to oscillations in the measured radiance that arise once the bright oxide crust forms on the surface and likely detach from the underlying substrate.
Assuming no significant modification of the surface during cooldown, the segments connecting markers E-F represent the temperature behavior of the oxidized surface emissivity after plasma exposure. 
In TiG5-D, all bands exhibit a consistent decreasing trend, with $ \epsstar $ declining from approximately 0.8 to 0.6 at $\SI{1000}{\kelvin}$. In TiG5-B, they remain constant or slightly increase with respect to the value measured at the end of the plasma exposure.

Additionally, emissivity in the $ \IRMband{0.95}{1.1} $ and $ \IRMband{0.6}{39} $ bands is compared with literature values reported for both virgin and oxidized Ti–6Al–4V surfaces. The present measurements were acquired at observation angles of approximately $\SI{35}{\degree}$ and $\SI{50}{\degree}$ relative to the surface normal, respectively; however, oxide emissivity is expected to exhibit significant angular dependence only at larger viewing angles \cite{Balat-Pichelin2020a}.
The data obtained in test TiG5-D show relatively good agreement with the virgin and oxidized normal spectral emissivity at $\SI{960}{\nm}$ ($\varepsilon_{960}^{\perp}$) reported by \citet{pagan2016}, although a more pronounced increase is observed above 1500~K. A similar trend is found in test TiG5-B during the early stages of the experiment (A–C), but substantial deviations arise following the temperature jump (C–D) and throughout the subsequent cooldown phase (E–F).
With regard to total emissivity, the present measurements begin only once surface oxidation has already initiated, and therefore do not capture the values of total normal emissivity ($\varepsilon_{\textrm{tot}}^{\perp}$) of the virgin surface reported by \citet{Balat-Pichelin2020a}. Nevertheless, prior to the temperature jump (B), both TiG5-D and TiG5-B approach values comparable to those of the pre-oxidized surface. While these emissivity levels are recovered after the jump in TiG5-D, TiG5-B instead stabilizes at intermediate values under steady-state conditions.

The present results are further compared with the total hemispherical emissivity ($\varepsilon\subsuprm{tot}{\cap}$) model proposed by \citet{Balat-Pichelin2020a}, which was derived from total angular radiance measurements during oxidation in dissociated oxygen flows. This model predicts an emissivity increase from approximately 0.25 below $\SI{1000}{\kelvin}$, corresponding to a non-oxidized surface, to about 0.8 in the temperature range between 1300 and $\SI{1874}{\kelvin}$.
Although the overall trend observed in the higher-temperature TiG5-D case is broadly consistent with the literature model, the in-situ measurements obtained in the present work reveal notable differences. In particular, the increase in $ \epsstar $ prior to the temperature jump (A–B) occurs at higher temperatures, namely between 1300 and $\SI{1500}{\kelvin}$, rather than within the 1100–$\SI{1300}{\kelvin}$ interval of the reference model. This shift may be associated with the higher heating rates \cite{rajabi2020} achieved in the present experiments.
Furthermore, while the TiG5-D measurements converge reasonably well toward the literature model, significant deviations are observed for TiG5-B, consistent with the distinct oxidation behavior exhibited during that test.

Overall, these findings emphasize the complex evolution of surface emissivity during high-enthalpy plasma wind tunnel oxidation. In particular, they show that 1) the observed non-monotonic behavior cannot be accurately captured using simplified temperature-dependent correlations alone, and 2) emissivity remains strongly influenced by the instantaneous oxidation state and surface morphology, as evidenced by the differing responses observed in TiG5-D and TiG5-B.

\subsection{Limitation of PWT experiments}
A key limitation of the PWT experiments presented in this work with respect to actual flight conditions is the inability to reproduce aerodynamic shear stresses and coupled motion effects such as tumbling. The brittle oxide layers observed during the experiments would likely experience cracking, spallation, or complete delamination under the aerodynamic loading encountered during a real atmospheric re-entry. Once detached, the underlying metallic surface becomes re-exposed to a highly oxidizing environment, effectively reinitiating the local oxidation process. As a result, the material response in flight is expected to involve a dynamic interplay between oxide growth and oxide removal, which cannot presently be reproduced in conventional PWT configurations.

Such peeling mechanisms may have important implications for spacecraft demise behavior. Repeated exposure of fresh material surfaces could enhance the effective oxidation rate compared to predictions based on uninterrupted oxide-scale growth. As a result, the emissivity evolution during flight may differ from that observed in ground-based experiments, owing to the continuous modification and renewal of the surface state. Oxide spallation may also contribute directly to mass loss and fragmentation processes, thereby affecting the overall survivability and breakup behavior of the object during re-entry.

Despite these limitations, PWT experiments remain a valuable tool for isolating and investigating fundamental oxidation and radiative phenomena under controlled, flight-representative high-enthalpy conditions, thus providing important insights into the transient coupling between oxidation, emissivity evolution, and material response.

%
\section{Conclusions}

Experiments were conducted in synthetic-air plasma on both Grade 5 (Ti-6Al-4V) and Grade 2 (commercially pure) titanium samples in order to reproduce the enthalpy levels and boundary-layer environments experienced by spacecraft tanks during hypersonic atmospheric re-entry. The primary objective was to obtain in situ emissivity measurements of actively oxidizing titanium surfaces under flow conditions more representative of atmospheric entry than those reported in the literature.

A sudden temperature jump was observed for both titanium grades at wall temperatures between 1500 and 1670 K, and was associated with the formation of a thick yellow oxide scale. The growth of this oxide layer was highly abrupt, often spreading over a large surface region within approximately 1 s. Post-test microscopy revealed complex and heterogeneous oxide morphologies, consisting of multiple layers with varying elemental compositions. Although the oxide crust generally remained attached after plasma exposure, it was brittle and readily delaminated from the solid substrate, whereas in regions that underwent melting and resolidification -- particularly for pure titanium -- it remained adherent and sticky.

The present results provide new insight into the dynamic evolution of emissivity during oxidation and melting across different spectral ranges, including the near-infrared, mid-infrared, and total emissivity bands. Emissivity increased progressively during oxidation, before undergoing a sudden decrease concurrent with the characteristic temperature jump. This transition is consistent with partial detachment of the oxide layer from the underlying substrate. Comparisons with literature data revealed discrepancies with in situ measurements obtained in this work. The delayed emissivity increase with temperature was attributed to the higher heating rates achieved during plasma wind tunnel tests, whereas its behavior following the temperature jump appeared to depend on the specific flow conditions, oxidation status, and proximity to the material melting point. Nevertheless, an important limitation of plasma wind tunnel experiments in extrapolating these observations to real flight conditions is the inability to reproduce aerodynamic shear stresses, which would likely promote repeated oxide-layer peeling and reformation.

Overall, these findings highlight the complex oxidation and emissivity behavior of titanium under entry-relevant conditions and emphasize the need for further investigation into the mechanisms governing the temperature jump, emissivity transition, and associated oxidation kinetics. The measured emissivity trends also carry important implications for space-debris demise modeling, suggesting that existing emissivity correlations may not adequately capture the transient behavior observed under more realistic re-entry environments.

%
\section*{Acknowledgments}
The experimental activities of this work were supported by the European Space Agency General Support Technology Program “Validation of Space Debris Demise Tools Using Plasma Wind Tunnel Testing and Numerical Tools” (contract n. 4000125437/18/NL/RA).
The research of A. Fagnani was funded by the Research Foundation--Flanders (dossier n. 1SB3221N). The authors would like to acknowledge the support of J.L. Freitas Monteiro and P. Collins
(Plasmatron facility operations), P. Laha and M. Raes (SEM/EDS data collection).

%
\appendix
\section{Validation of two-color pyrometry measurements}
\label{sec:appendix_TCP}

The emissivity measurements presented in this work are based on surface temperatures obtained via two-color ratio pyrometry. The reliability of this technique depends on its sensitivity to both line-of-sight and scattered radiation, as well as potential deviations from grey-body surface behavior within the probed spectral range. These effects are examined in detail herein.

\begin{figure}[h]
	\centering
	\subfigure[]
	{\includegraphics[width=0.99\linewidth, trim={0cm, 0.5cm, 0cm, 0cm}, clip]{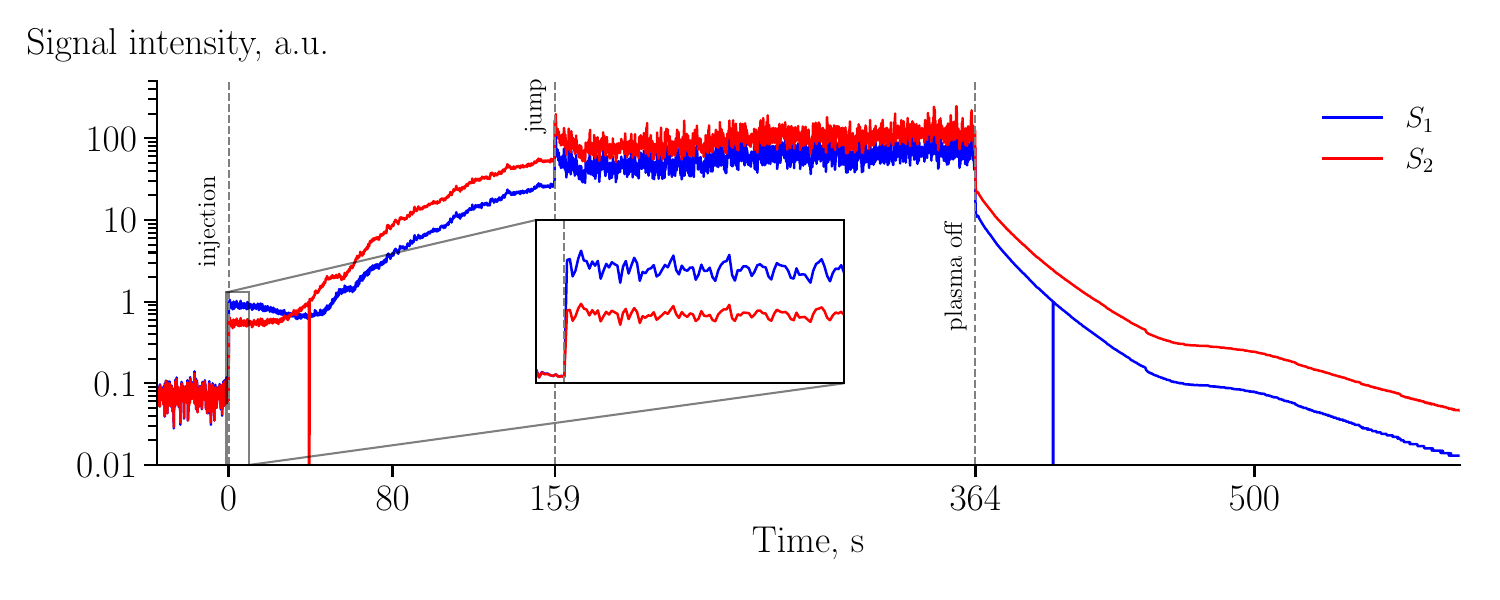}	\label{fig:figureA10a}}
	\subfigure[]
	{\includegraphics[width=0.99\linewidth, trim={0cm, 0.5cm, 0cm, 0cm}, clip]{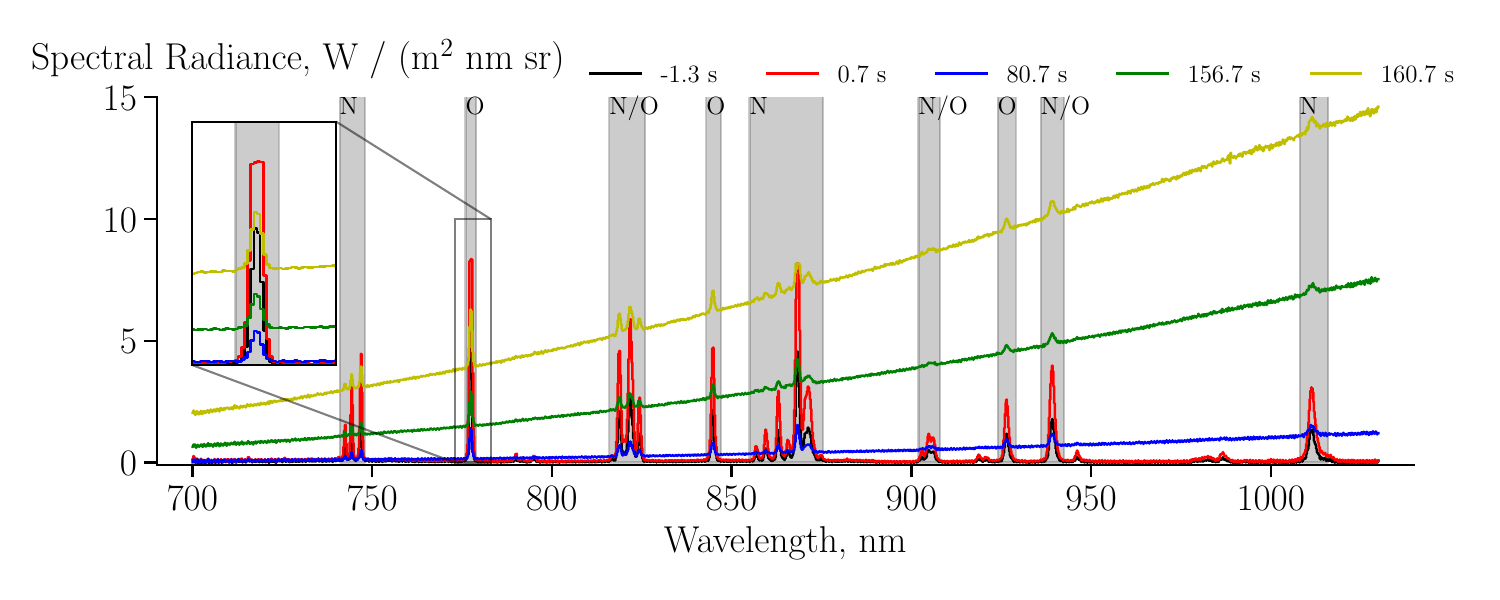}	\label{fig:figureA10b}}\\
	\subfigure[]
	{\includegraphics[width=0.45\linewidth, trim={0cm, 0.5cm, 0cm, 0cm}, clip]{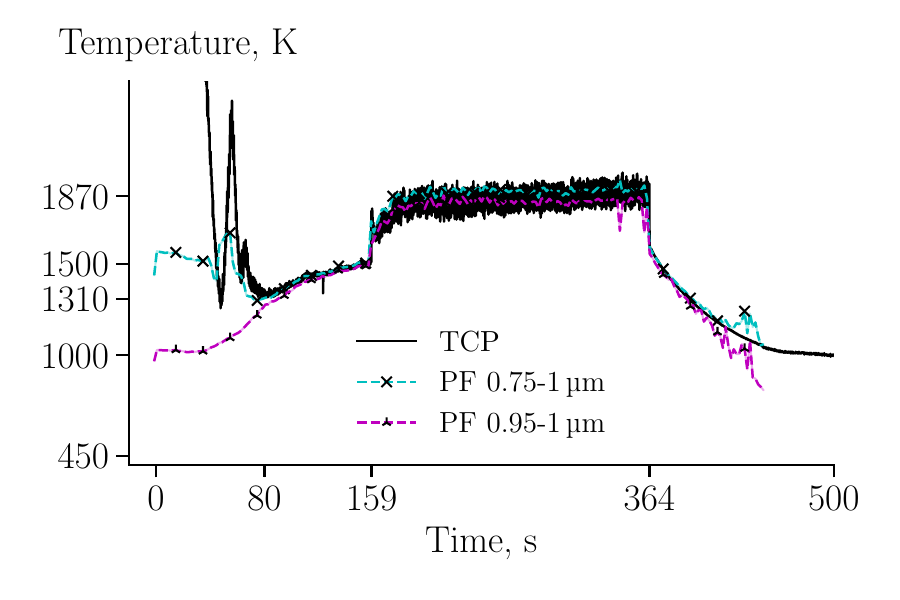}	\label{fig:figureA10c}}
	\subfigure[]
	{\includegraphics[width=0.45\linewidth, trim={0cm, 0.5cm, 0cm, 0cm}, clip]{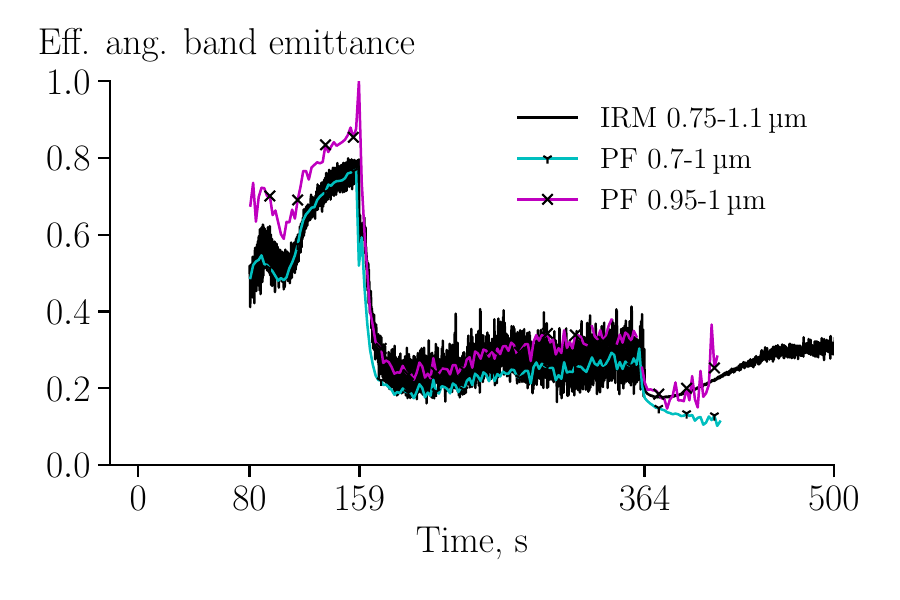}	\label{fig:figureA10d}}
	\caption[Validation of the TCP measurements for titanium.]{TiG5-B: (a) Positive jump in the raw two-color pyrometer signals ($S_1$ and $S_2$) at injection suggest significant scattered light emitted by the plasma torch, confirmed by the increase in the detected atomic line intensities by surface spectrometry (black to red line) (b). (c) TCP temperature agrees with Planck fitting method after 80~s, after which surface temperature measurement is considered reliable. Agreement after jump phenomenon further corroborates measurements. (d) Measured band emissivity $\epsstar$ agrees with Plack Fitting (PF) within $\sim10\%$ uncertainty. }
	
\end{figure}

First, the correction methodology for line-of-sight plasma radiation developed in Ref.~\cite{fagnani2024a} exhibits limitations when applied to curved, reflective surfaces. As shown in Fig.~\ref{fig:figureA10a}, the raw signal intensities from test TiG5-B increase immediately upon injection, in contrast to the behavior typically observed for high-emissivity flat samples, where a decrease in recorded intensity is expected. This discrepancy is attributed to the scattering of the intense plasma torch emission by the sample surface, a phenomenon also evident in the camera images at injection, e.g., in Fig.~\ref{fig:figure9_7}(a).

Since emissivity evolves significantly during plasma exposure, the use of early-time signals for correction is not appropriate, as the scattered contribution is not constant. Simultaneous emission spectroscopy of the surface in Fig.~\ref{fig:figureA10b}, employing the method described in Ref.~\cite{fagnani2024a}, supports this interpretation, showing a marked increase in atomic emission lines from O and N immediately after injection (between $-1.3$ and $0.7~\si{\second}$). This is attributed to the reflected signal from inside the torch, where their concentration is more abundant and temperatures are considerably higher.
As both surface temperature and emissivity increase significantly at later times, the scattered contribution is expected to become less relevant. Indeed, the broadband, Planck-like emission associated with the heated surface becomes detectable after approximately $80~\si{\second}$ and increases sharply during the observed temperature jump (156.7–160.7~s). Correspondingly, the reflected contribution from the atomic lines becomes less relevant as the surface emissivity increases.

To further validate the TCP-derived surface temperatures, the measured surface spectral radiance $ L\subrm{meas} $ is fitted with a Planck-like distribution $ \varepsilon\subrm{PF}  \cdot L\subrm{BB}(T) $, where $ L\subrm{BB}(T)  $ is the blackbody spectral radiance for a surface temperature $T$. The fitting is performed over the wavelength bands $ \IRMband{0.75}{1} $ and $ \IRMband{0.95}{1} $, corresponding to the spectral sensitivity of the MR1SB two-color pyrometer used for TCP measurements, while assuming a uniform emissivity $ \varepsilon\subrm{PF} $ within each band. Spectral regions contaminated by strong atomic emission lines (indicated as grey areas in Fig.~\ref{fig:figureA10b}) were excluded to minimize interference from scattered gas radiation.

Figure~\ref{fig:figureA10c} shows that the temperatures obtained from TCP and Planck fitting agree within 1\% after $80~\si{\second}$, which is therefore identified as the threshold beyond which TCP measurements can be considered reliable, as presented in Fig.~\ref{fig:figure7}. A similar procedure was carried out for TiG5-D and TiG2-A. This agreement also supports the validity of the temperature values measured after the abrupt transition, despite potential variations in spectral emissivity between the two pyrometer bands. Furthermore, as illustrated in Fig.~\ref{fig:figureA10d}, the consistency between band emissivity values derived from the IRM method and those obtained via Planck fitting provides additional validation of the in situ measurement approach, within an uncertainty of approximately 10\% at the investigated temperatures.

%

\bibliographystyle{elsarticle-num-names} 
\bibliography{biblio}

\end{document}